\documentclass[bibyear]{aa} % if the references are not structured 
\usepackage{graphicx}
\usepackage{txfonts}
\usepackage{color}{}

\begin{document}

   \title{\textit{Herschel} observations of the nebula M1-67 around\\
                the Wolf-Rayet star WR 124 
               \thanks{\textit{Herschel} is an ESA space observatory 
                with science instruments provided by European-led Principal 
                Investigator consortia and with important participation from
                NASA.}$^{,}$
                \thanks{Based in part on observations collected at the European
                Southern Observatory, La Silla, Chile}}

   \author{C. Vamvatira-Nakou\inst{1}
          \and D. Hutsem\'{e}kers\inst{1}\fnmsep\thanks{Senior Research 
               Associate FNRS}
          \and P. Royer\inst{2} 
           \and C. Waelkens\inst{2} 
          \and  M. A. T. Groenewegen\inst{3} 
          \and  M. J. Barlow\inst{4}
          }

   \institute{Institut d'Astrophysique et de G\'{e}ophysique, Universit\'{e} de Li\`ege, Quartier Agora, All\'{e}e du 
              6 ao\^ut, 19C - B\^at. B5c, B-4000 Li\`ege (Sart-Tilman), Belgium \\
              \email{C.VamvatiraNakou@alumni.ulg.ac.be}
           \and Instituut voor Sterrenkunde, KU Leuven, Celestijnenlaan 200D,
                   Bus 2401, B-3001 Leuven, Belgium
           \and Koninklijke Sterrenwacht van Belgi\"e, Ringlaan 3, B-1180 Brussels, Belgium
           \and Department of Physics and Astronomy, University College London, Gower Street, London WC1E 6BT, UK
             }

%   \date{Received 2014; accepted 2015}
\date{Preprint online version: February 10, 2016}

% \abstract{}{}{}{}{} 
% 5 {} token are mandatory

\abstract{
Infrared {\it Herschel} imaging and spectroscopic observations
of the nebula M1-67 around the Wolf-Rayet star WR 124 have been obtained along
with optical imaging observations. The infrared images reveal a clumpy dusty
nebula that extends up to 1 pc. The comparison with the optical images shows
that the ionized gas nebula coincides with the dust nebula, the dust and
the gas being mixed together. A
photodissociation region is revealed from the infrared spectroscopic
analysis. The analysis of the infrared spectrum of the nebula, where
forbidden emission lines of ionized elements were detected, showed
that the nebula consists of mildly processed material with the calculated abundance
number ratios being N/O = 1.0  $\pm$ 0.5 and C/O = 0.46 $\pm$ 0.27. Based on a radiative
transfer model, the dust mass of the nebula was estimated to be 0.22 M$_{\odot}$
with a population of large grains being necessary to reproduce the observations.
The comparison of the mass-loss rate and the abundance ratios to theoretical
models of stellar evolution led to the conclusion that the nebular
ejection took place during a RSG/YSG evolutionary phase of a central star with
an initial mass of 32 M$_{\odot}$.
}

   \keywords{circumstellar matter --
             Stars: massive --
             Stars: mass-loss --
             Stars: Wolf-Rayet --
             Stars: individual: WR 124 }

   \authorrunning{C. Vamvatira-Nakou et al.}
   \titlerunning{\textit{Herschel} observations of the nebula M1-67}

   \maketitle

%
%________________________________________________________________
\section{Introduction}
\label{sec:introduction}

Wolf-Rayet (WR) stars represent an intermediate phase in the late evolution
of O-type massive stars with an initial mass $\ge 30\ \mathrm{M}_{\odot}$
(Maeder \& Meynet \cite{maed10}). The star loses a significant fraction of
its mass through the stellar wind and/or through episodes of extreme
mass loss during a red supergiant (RSG) or luminous blue variable
(LBV) evolutionary phase so that the outer layers are removed, leaving a
bare core that becomes a WR. If the WR is in a close binary system the lower
limit for the initial mass is not as robust (Crowther \cite{crow07}), and in
this case the H-rich envelope is lost through a Roche lobe overflow. WR stars
are characterized by strong broad emission lines in the optical region
due to stellar winds. They are divided into two groups: the WN subtypes that
show strong lines of He and N and the WC and WO subtypes that show strong He,
C, and O in their spectra (Crowther \cite{crow07}).

One third of the galactic WR stars are observed to have an associated nebula
at optical wavelengths (Marston \cite{mar97}). Various morphologies have
been observed around Galactic WR stars (Chu et al. \cite{chu83}) and WR
stars in the Magellanic Clouds (Dopita et al. \cite{dop94}). The ring
nebulae around WR stars are thought to contain material that has been
ejected in a previous evolutionary phase of the star, a LBV or a RSG
phase (Crowther \cite{crow07}). Besides this, ejected nebula have been observed
around LBV stars (Hutsem\'{e}kers \cite{hut94}; Nota et al. \cite{nota95}).
Consequently, the study of the nebulae around WR stars and in general around
evolved massive stars is crucial for understanding the evolution of these
stars, in particular, through the study of their mass-loss history.

The nebula M1-67 surrounds the Galactic star WR 124 (209 BAC, also known
as Merrill's star), which has a WN 8 spectral type (Smith and Aller
\cite{smi69}). It was discovered by Minkowski (\cite{minkov46}), who
suggested that it might be a planetary nebula (PN). Then, Sharpless
(\cite{sharp59}) classified it as an \ion{H}{ii} region. Merrill
(\cite{mer38}) measured a high heliocentric velocity of
$+200$ km s$^{-1}$ for WR 124 that was confirmed by Bertola (\cite{bert64}).
The nebula M1-67 was later found to have a similar heliocentric velocity
(Perek and Kohoutek \cite{per67}; Cohen and Barlow \cite{coh75}; Pismis
and Recillas-Cruz \cite{pism79}). It was first suggested to be a WR ring
nebula in the study of Cohen and Barlow (\cite{coh75}) based on infrared
observations.

\begin{figure*}[!]
\resizebox{\hsize}{!}{\includegraphics*{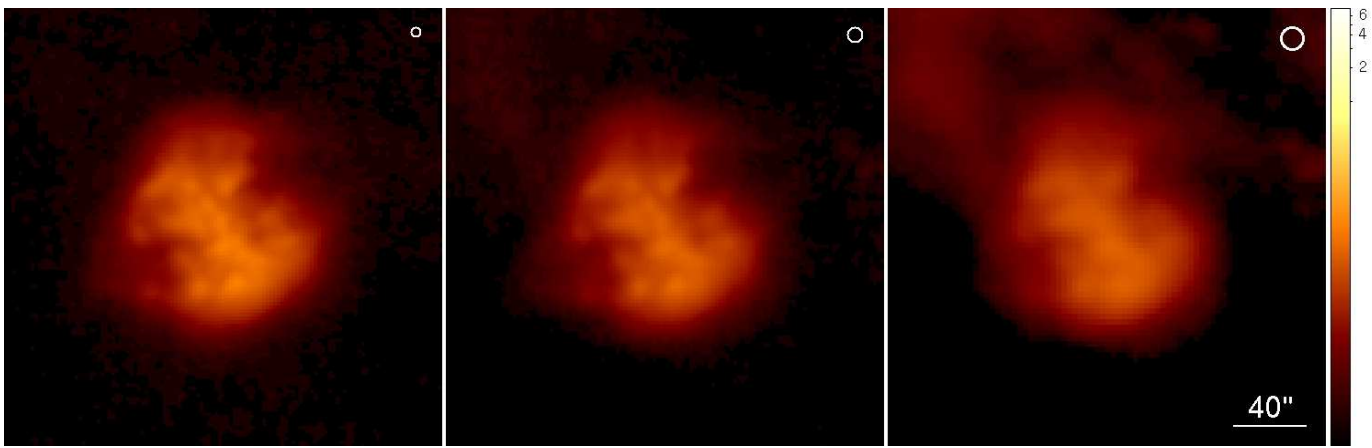}}\\%
\caption{PACS images of the nebula M1-67 at 70\
\mbox{$\mu$m}, 100\ \mbox{$\mu$m}, and 160\ \mbox{$\mu$m}, from left to
right. The size of each image is 4$\arcmin\times4\arcmin$.
The scale on the right corresponds to the surface brightness (arbitrary
units) and is logarithmic to better show the faint emission. The
white circles in the upper right corner of each image represent the PSF size.
North is up and east to the left.}
\label{m1-67_PACS}
\end{figure*}

Based on the analysis of high-resolution spectra of the nebula M1-67, Solf
and Carsenty (\cite{solf82}) argued that the nebular material originates
in the central WR star. This material may have been lost in an earlier
evolutionary stage of the central star and swept up by the strong
stellar wind. The study of Esteban et al. (\cite{est91}) confirmed the stellar
origin of the nebular material. They drew this conclusion based on the
abundances of N (enhanced by a factor of 4-7.5), O (depleted by a factor of
5-7.5) and S (normal \ion{H}{ii} region abundance), and concluded that M1-67
is a WR ring nebula. Though the nebula is spatially non-uniform and has a
clumpy appearance with knots of emission, they also found that there is a
homogeneity in excitation conditions throughout the nebula, since the knots have
similar spectra.

Analyzing optical coronagraphic data, Nota et al. (\cite{nota95_m})
reported observing an axisymmetrical expansion in M1-67 for the first time.
Sirianni et al. (\cite{sir98}) performed similar
observations with higher resolution. Their radial velocity data
revealed two different motions in the circumstellar environment of
WR 124: the expansion of a spherical shell and a bipolar outflow that
both may have been formed during the post-main sequence evolution of the
central star when it was a LBV. Deep Hubble Space Telescope (HST) imaging
(Grosdidier et al. \cite{gro98}) revealed details of the fragmentation of
the nebula, which appeared composed of filamentary structures that
extend around the central star and many, mostly unresolved, clumps. These
data showed no clear evidence of a bipolar or axisymmetrical structure.
The authors argued that M1-67 could be the result of a clumpy wind during
the LBV evolutionary phase of the central star.

Van der Sluys and Lamers (\cite{sluy03}) studied the dynamics of the M1-67
nebula. They show that it interacts with the interstellar medium
so that a parabolic-like bow shock is formed. Because the star is moving
away from us with a velocity of about $180$ km s$^{-1}$, we see the hollow
bow shock from behind (see also Solf and Carsenty \cite{solf82}).
They conclude that the M1-67 nebula is due to
multiple outbursts during a previous LBV phase of the central star flowing
along the bow-shock surface. Cichowolski et al. (\cite{cicho08}) analyzed
high resolution radio data of the circumstellar environment of WR 124. They
reporte the presence of two large \ion{H}{i} cavities that are parts of the
same bow shock structure. The radio morphology of the nebula at 8.5 GHz is
very similar to the optical one.
Based on integral field spectroscopic data and considering theoretical evolutionary
models, Fern\'{a}ndez-Mart\'{i}n et al. (\cite{fern13}) propose
that the central star has recently entered the WR phase and that the nebula
is essentially composed of material that was ejected during a LBV phase of
the star.

Concerning the distance to M1-67, several studies tried to determine it, but
given the peculiar velocity of the nebula, they had to assume the nature of
the central star and its nebula. Cohen and Barlow (\cite{coh75}) calculated
a distance of $4.33$ kpc, arguing that the central star is a Population
I WN8 star and using the absolute magnitude calibration of Smith (\cite{smi73}),
a value consistent with the observed extinction. Pismis and Recillas-Cruz
(\cite{pism79}) estimated a distance of about $4.5$ kpc, also assuming a
Population I WR star of spectral type WN8 but using a different and more
recent absolute magnitude calibration. Later on, Crawford and Barlow (\cite{craw91})
analyzed the interstellar Na I D$_2$ absorption spectrum of the star and found
a distance of $4-5$ kpc that excluded a PN nature. Recently, Marchenko et
al. (\cite{march10}) have calculated a geometric distance of $3.35$ kpc with an
estimated uncertainty of $20$\%, based on two-epoch imaging data that provide
an expansion parallax. Since this latest distance calculation is assumption-free
as stressed by the authors, we adopt it for our study, along with the properties
of the star and the wind that were derived based on it in Marchenko et al.
(\cite{march10}).

In the present paper we analyze and discuss the infrared images and the spectrum
of the nebula M1-67 taken with the PACS (Photodetector
Array Camera and Spectrometer, Poglitsch et al. \cite{poglitsch}),
and SPIRE (Spectral and Photometric Imaging Receiver, Griffin et al. \cite{grif10})
instruments onboard the $\textit{Herschel}$ Space Observatory (Pilbratt
et al. \cite{pilbratt}). We present these observations and the data reduction
procedure in Sect.~\ref{sec:observations and data reduction}. Using these
observations, we then describe the nebular morphology in
Sect.~\ref{sec:morphology of the nebula}. In Sect.~\ref{sec:dust
continuum emission} we perform the dust continuum emission model,
while the analysis of the emission line spectrum follows in
Sect.~\ref{sec:emission line spectrum}. In Sect.~\ref{sec:discussion} a
general discussion is presented and finally in Sect.~\ref{sec:conclusions}
the conclusions of this work are given.

%______________________________________________________________
\section{Observations and data reduction}
\label{sec:observations and data reduction}

\subsection{Infrared observations}

\begin{figure*}[!]
\resizebox{\hsize}{!}{\includegraphics*{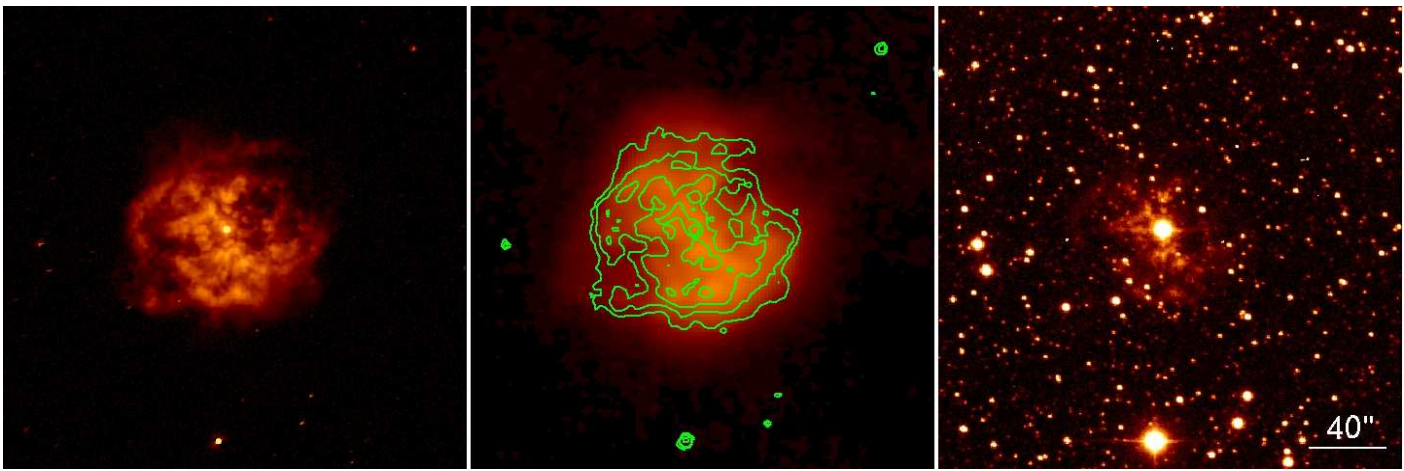}}\\%
\caption{Left:View of the optical H$\alpha$+[N{\sc ii}] emission from the
nebula M1-67. Center: Contour image of the H$\alpha$+[N{\sc ii}] emission from
the nebula (green lines) superposed on the PACS infrared image of the nebula at
70\ \mbox{$\mu$m} (shown also in Fig.~\ref{m1-67_PACS} at the same scale). Right:
View of the nebula in the continuum filter. This continuum image has been
subtracted from the image obtained in the H$\alpha$+[N{\sc ii}] filter to create
the image seen in the leftmost panel and to emphasize the ionized gas emission.
The size of each image is 4$\arcmin\times4\arcmin$ and the scale is logarithmic
to better show the faint emission. North is up and east to the left.}
\label{m1-67_Ha_IR_cont}
\end{figure*}

The infrared observations of the M1-67 nebula were carried out in the framework
of the \textit{\emph{Mass-loss of Evolved StarS} (MESS)} Guaranteed Time Key Program
(Groenewegen et al. \cite{groenewegen}), including PACS and
SPIRE imaging and PACS spectroscopy.

The PACS imaging observations were carried out on April 8, 2010,
which corresponds to \textit{Herschel}'s observational day (OD) 329.
The scan map observing mode was used, in which the telescope slews
at constant speed ($20''/\mbox{s}$) along parallel lines so as
to cover the required area of the sky. Two orthogonal scan maps
were obtained for each filter and finally our data set consists
of maps at 70, 100, and 160\ \mbox{$\mu$m}. The observation
identification numbers (obsID)s of the four scans are 1342194080,
1342194081, 1342194082, and 1342194083 with a duration of 2622 s
each. The Herschel Interactive Processing Environment (HIPE) package (Ott
\cite{ott10}) was used for the data reduction up to level 1.
Subsequently, the Scanamorphos software (Roussel \cite{rous12})
was used to further reduce and combine the data. In the final maps,
the pixel size is 2$\arcsec$ in the blue (70, 100 $\mu$m) channel
and 3$\arcsec$ in the red (160 $\mu$m) channel. The $\textit{Herschel}$
PACS point-spread function (PSF) full widths at half maximum (FWHMs)
are 5$\farcs$2, 7$\farcs$7, and 12$\arcsec$ at 70\ \mbox{$\mu$m},
100\ \mbox{$\mu$m}, and 160\ \mbox{$\mu$m}, respectively.

The SPIRE imaging observations were carried out on September 21,
2010 (OD 495). The large map observing mode was used. In this mode
the telescope slews at constant speed (nominal speed:
$30''/\mbox{s}$) along parallel lines so as to scan the required
sky area. The cross scan pointing mode was selected so as to obtain
two orthogonal scans during a single observation. Our dataset consists
of maps at 250, 350, and 500\ \mbox{$\mu$m}. The obsID is 1342204949
with a duration of 911 s. The data were retrieved from the archive,
processed up to level 2. The surface brightness of the three maps
was transformed from Jy/sr to Jy/pixel with the help of HIPE.
The pixel size is 6$\arcsec$, 10$\arcsec$, and 14$\arcsec$ at
250\ \mbox{$\mu$m}, 350\ \mbox{$\mu$m}, and 500\ \mbox{$\mu$m},
respectively. The SPIRE PSF FWHMs are 18$\arcsec$, 25$\arcsec$ and
37$\arcsec$ at 250\ \mbox{$\mu$m}, 350\ \mbox{$\mu$m}, and
500\ \mbox{$\mu$m}, respectively.

The PACS spectroscopic observations of the M1-67 nebula were
carried out on May 4, 2011 (OD 720). The PACS integral-field
spectrometer covers the wavelength range from 52\ \mbox{$\mu$m}
to 220\ \mbox{$\mu$m} in two channels that operate simultaneously
in the blue band, 52-98\ \mbox{$\mu$m}, and the red band, 102-220\ 
\mbox{$\mu$m} with a resolving power of $\lambda/\delta\lambda
\sim 940-5500$ depending on the wavelength. Simultaneous imaging
of a $47\arcsec \times 47\arcsec$ field of view is provided,
resolved in $5\times5$ square spatial pixels (i.e., spaxels). Then,
the two-dimensional field-of-view is rearranged via an image slicer
along a $1\times25$ pixels entrance slit for the grating. The chopped
line scan observing mode was used for these observations. In this mode,
instead of having a complete coverage between 52\ \mbox{$\mu$m} and
220\ \mbox{$\mu$m}, the observations are done in specific, previously
chosen, short wavebands where it is possible to detect a nebular
emission line. The background spectrum was obtained through chopping
and nodding. The two obsIDs are 1342220598 and 1342220599. For
the spectroscopic data reduction we followed the standard steps
using HIPE.

\subsection{Visible observations}

The optical images of the M1-67 nebulae were obtained on April 6,
1995, with the 3.6-m telescope at the ESO, La Silla, Chile.
A series of short (1 s - 10 s) and long (30
s - 60 s) exposures were secured in a H$\alpha$+$[\ion{N}{ii}]$ filter
($\lambda_{\rm c}$ = 6560.5 \AA ; {\sc FWHM} = 62.2 \AA ) and in a
continuum filter just redwards ($\lambda_{\rm c}$ = 6644.7 \AA; 
{\sc FWHM} = 61.0 \AA ). The frames were bias-corrected and flat-fielded.
The night was photometric and the seeing around 1$\farcs$2. The CCD
pixel size was 0$\farcs$605 on the sky. To properly calibrate the images,
three spectrophotometric standard stars and three PN with
known H$\alpha$ flux were observed.

%__________________________________________________________________
\section{Morphology of the nebula M1-67}
\label{sec:morphology of the nebula}

\begin{figure}[!]
\resizebox{\hsize}{!}{\includegraphics*{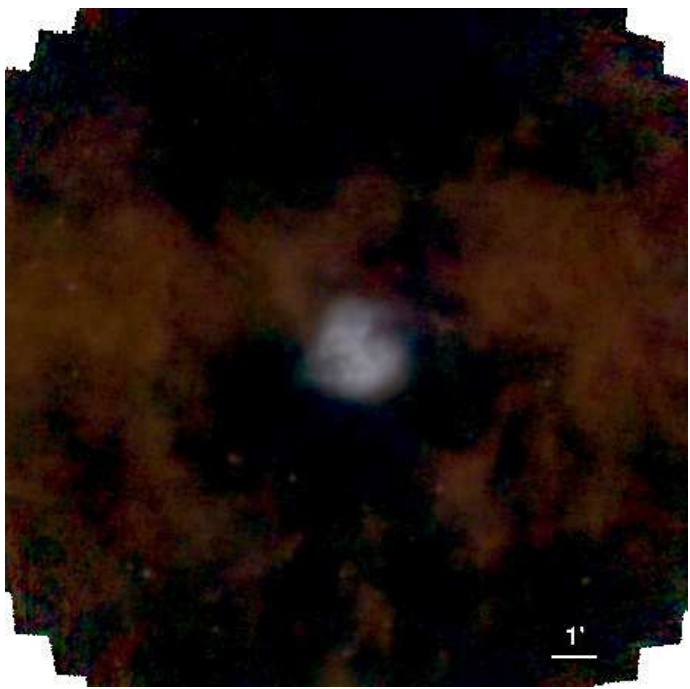}}
\caption{Three-color (70 \ \mbox{$\mu$m} in blue, 100 \ \mbox{$\mu$m}
in green, and 160 \ \mbox{$\mu$m} in red) image of the nebula M1-67.
The nebula appears located inside a cavity in the interstellar
medium. The scale is logarithmic for all three colors.
The size of the image is 15$\arcmin\times15\arcmin$. North is
up and east to the left.}
\label{m1-67_IR_RGB}
\end{figure}

The PACS images of the nebula M1-67 around the WR 124 star in the three
infrared filters, 70\ \mbox{$\mu$m}, 100\ \mbox{$\mu$m}, and 160\ \mbox{$\mu$m},
are shown in Fig.~\ref{m1-67_PACS}\footnote{The SPIRE images are not presented
here because the nebula is barely resolved because of the low resolution.}. The
optical H$\alpha$+$[\ion{N}{ii}]$ image of the nebula M1-67 is illustrated in
Fig.~\ref{m1-67_Ha_IR_cont}.
To emphasize the H$\alpha$+$[\ion{N}{ii}]$ ionized gas emission, the redward
continuum image has been subtracted, after correcting for the position offsets
and for the different filter transmissions, using field stars. It should
be mentioned that the nebula is clearly detected in the $\lambda$ 6644.7 \AA\
continuum filter (right panel of Fig.~\ref{m1-67_Ha_IR_cont}), indicating significant dust
scattering. In the central panel of Fig.~\ref{m1-67_Ha_IR_cont}, the contour
image of the H$\alpha$+$[\ion{N}{ii}]$ emission is illustrated, superimposed
on the infrared image of the nebula at 70\ \mbox{$\mu$m}. A three color image
of the nebula M1-67 and its environment is shown in Fig.~\ref{m1-67_IR_RGB}.

The infrared images reveal a dust nebula with a non-uniform brightness
that has a complex morphology.
It appears to be clumpy but the clumps are not resolved in these observations.
Although the brightest parts of the nebula appear distributed along an elongated
structure, it is difficult to infer from the infrared images that the
nebula is bipolar as suggested by Fern\'{a}ndez-Mart\'{i}n et al. (\cite{fern13}).
Moreover, a fainter structure, with a roughly spherical shape, seems to surround the
bright nebulosities. This faint spherical structure was also observed
in the Spitzer MIPS 24\ \mbox{$\mu$m} image (Gvaramadze et al. \cite{gvar10})
discussed in Fern\'{a}ndez-Mart\'{i}n et al. (\cite{fern13}). The global infrared
morphology of M1-67 therefore seems to be essentially spherical,
although strongly inhomogeneous,
with an average radius of about 60$\arcsec$ that corresponds to $1$ pc at the
adopted distance of $3.35$ kpc.
The nebula seems to be located inside an empty cavity (Fig.~\ref{m1-67_IR_RGB})
that may have formed in a previous evolutionary phase of the central
star.

The H$\alpha$+$[\ion{N}{ii}]$ view of the nebula is very similar to the infrared one.
Its general morphology is the same, but with more details because the optical images
have a higher resolution (Fig.~\ref{m1-67_Ha_IR_cont}). The dust nebula extends
slightly further out the gas nebula. The ionized gas nebula is clumpy with
a very complex structure of filaments. The contour image of the optical emission
superimposed on the 70\ \mbox{$\mu$m} image shows that the brightest regions
(clumps) of the ionized gas nebula coincide with the bright regions of the dust
nebula. An average angular radius of $\sim$ 55$\arcsec$, which corresponds to
$0.9$ pc at a distance of $3.35$ kpc, can be defined for the ionized gas nebula,
in agreement with the measurements of Grosdidier et al. (\cite{gro98}). Solf \&
Carsenty (\cite{solf82}) and Sirianni et al. (\cite{sir98}) had measured a smaller
radius of about 45$\arcsec$ without detecting the faint outer component. The similarity
between the optical and the infrared view implies that the gas is mixed with the
dust in the nebula M1-67.

%__________________________________________________________________
\section{Dust continuum emission}
\label{sec:dust continuum emission}

By performing aperture photometry on the PACS and SPIRE images, we derived
integrated flux densities for the nebula M1-67. Data from the archive of
the Infrared Astronomical Satellite (IRAS) mission (Neugebauer et al. \cite{neug84})
were also used to extend the spectral energy distribution (SED) to shorter wavelengths.
To all the flux densities derived from these data, we applied photometric color
corrections in order to make the conversion of the monochromatic flux densities,
which refer to a constant energy spectrum, to the true object flux
densities at the photometric reference wavelengths of each instrument.

For the IRAS data, the flux density ratios were used to derive the color
temperature, and the corresponding correction factor was used to correct the
flux densities (Beichman et al. \cite{beichman}). The ratio R (12,25) corresponds
to a temperature of $140$ K, while the ratios R (25,60) and R (60,100) correspond
to a temperature of $100$ K.
Consequently, we decided to calculate the correction for the flux densities
at 12 $\mu$m and 25 $\mu$m using both temperatures and then to consider the average
of the two corrected flux densities at each wavelength, counting the
difference between them in the error. The flux densities at 60 $\mu$m and
100 $\mu$m were corrected using the correction factors that correspond to
a temperature of $100$ K. For the \textit{Herschel}-PACS data color correction,
we fit a black body (BB), also using the 25 $\mu$m IRAS observation to have a
data point before the maximum of the BB curve. The color correction
factors that correspond to a temperature of $100$ K derived from this fit were then
used to correct the flux densities (M\"{u}ller et al. \cite{muller}). For the
\textit{Herschel}-SPIRE data color correction, we followed the instructions
given in SPIRE Handbook \footnote{http://herschel.esac.esa.int/Docs
/SPIRE/spire\textunderscore handbook.pdf}. 

\begin{figure}[!]
\resizebox{\hsize}{!}{\includegraphics*{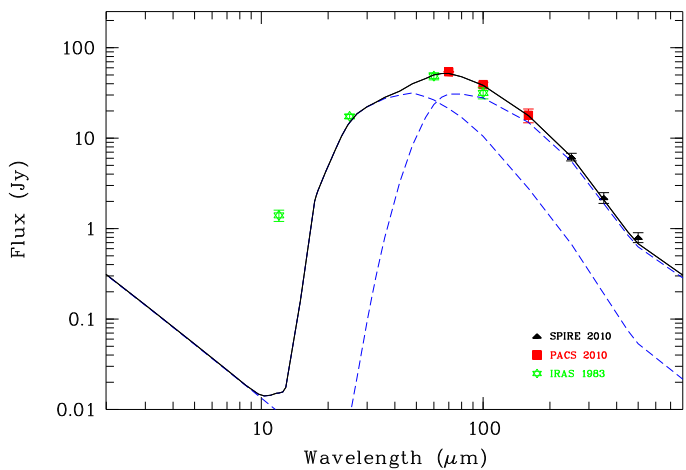}}
\caption{Infrared SED of the nebula M1-67 around
the Wolf-Rayet star WR 124 from
color-corrected photometric measurements from IRAS and \textit{Herschel}
(PACS and SPIRE) space missions. The best result of the 2-Dust model fitting is
illustrated (solid line). It was achieved by considering two populations of dust grains
(dashed lines), a population of small grains with radii from $0.005$ to $0.1$ $\mu$m,
responsible for the emission at $\lambda < 40 \mu$m and a population of large grains with
radii from $2$ to $10$ $\mu$m, responsible for the infrared SED at $\lambda > 70 \mu$m.
Data at $\lambda < $ 20 $\mu$m are not considered in the fit.}
\label{m1-67_2dust}
\end{figure}

Table~\ref{table:1} presents the color-corrected flux density measurements from
IRAS and \textit{Herschel} space missions. These data are used to construct
the infrared SED of the nebula M1-67 shown in Fig.~\ref{m1-67_2dust}.

\begin{table}[h]
\caption{Color-corrected flux densities of the nebula M1-67.}
\label{table:1}
\centering
\begin{tabular}{l c c c c c}
\hline\hline
Spacecraft-instrument &    Date & $\lambda$  & $F_{\nu}$ & Error  \\
                      &         & ($\mu$m)   & (Jy)      & (Jy)   \\
\hline\hline
IRAS\tablefootmark{a} &   1983  &   12       & 1.4   & 0.2   \\
                      &         &   25       & 17.4  & 0.9   \\
                      &         &   60       & 48.5  & 4.4   \\
                      &         &   100      & 31.6  & 4.4   \\
\hline
Herschel-PACS
\tablefootmark{b}     &  2010   &   70       & 54.2  & 0.5   \\
                      &         &   100      & 39.1  & 2.4   \\
                      &         &   160      & 17.9  & 3.1   \\
Herschel-SPIRE
\tablefootmark{b}     &  2010   &   250      & 6.2   & 0.6   \\
                      &         &   350      & 2.2   & 0.3   \\
                      &         &   500      & 0.8   & 0.1   \\
\hline
\end{tabular}
\tablefoot{Data from:
\tablefoottext{a}{Moshir et al. \cite{mosh92}.}
\tablefoottext{b}{This work.}
}
\end{table}

To model the dust nebula around WR 124 and calculate the temperature and the
mass of the dust we used the 2-Dust code (Ueta and Meixner \cite{uet03}), a
publicly available two-dimensional radiative transfer code that can be
supplied with complex axisymmetric density distributions as well as with
various dust grain density distributions and optical properties.
The photometric data of Table~\ref{table:1} are used for fitting the dust model.

The first step for the dust modeling is to constrain the nebular geometry
revealed through the optical and the infrared images. As mentioned in
Sect.~\ref{sec:introduction}, according to previous studies there are
two different scenarios for the nebula M1-67: a) a bipolar
morphology that was suggested by Sirianni et al. (\cite{sir98}),
firstly reported by Nota et al. (\cite{nota95_m}), but that was not confirmed
by Grosdidier et al. (\cite{gro98}, \cite{gro01}); and b) a bow shock
model that was suggested by Van der Sluys and Lamers (\cite{sluy03}) after a
detailed analysis of radial velocities, firstly mentioned by Solf and
Carsenty (\cite{solf82}) and later by Grosdidier et al.
(\cite{gro99}), and that was confirmed by Cichowolski
et al. (\cite{cicho08}) and Marchenko et al. (\cite{march10})
(see Solf and Carsenty \cite{solf82} for a sketch).

The bow shock scenario is the most plausible because the nebula M1-67 surrounds a
massive runaway star that moves with a velocity that is much higher than the velocity of the
interstellar medium (Solf and Carsenty \cite{solf82}; Van der Sluys and Lamers \cite{sluy03}). Also,
the optical and infrared view of the nebula does not show evidence for a clear
bipolar morphology (Sect.~\ref{sec:morphology of the nebula}). In this
case, the standoff distance of the bow shock, d$_s$, is about $0.65$ pc using the
parameters given in Marchenko et al. (\cite{march10}) and the stellar
velocity of $200$ km/s determined by Cichowolski et al. (\cite{cicho08}).
This distance is equivalent to 40$\arcsec$ at $3.35$ kpc. It is smaller than the measured
radius of the nebula in the infrared, which is 60$\arcsec$ or $1$ pc (Sect.~\ref{sec:morphology
of the nebula}), so the star is decentered with respect to the nebula but not much.
Consequently, to model the dust, we make the approximation that the nebula is
spherical with $r_{\mathrm{in}}$ = 40$\arcsec$ and  $r_{\mathrm{out}} = 60\arcsec$.

For the parameters of the central star WR 124 we adopted the distance
D = $3.4$ kpc, the luminosity $\log L/L_{\odot} = 5.18$, and the temperature
$T_{\mathrm{eff}} = 35800$ K, which were estimated by Marchenko et al.
(\cite{march10}). We considered two populations of dust grains
that have same composition but different sizes, because the infrared SED
of M1-67 (Fig.~ \ref{m1-67_2dust}) is too broad to be reproduced with only
one population as was the case for the nebula around the LBV AG
Car (Vamvatira-Nakou et al. \cite{vamv15}). For each of the two
populations of dust grains, we assumed the size distribution of Mathis et
al. (\cite{mat77}): $n(\alpha) \propto \alpha^{-3.5}$ with $\alpha_{\rm min} < \alpha < \alpha_{\rm max}$,
$\alpha$ being the grain radius. By varying $\alpha_{\rm min}$ (or $\alpha_{\rm max}$) and
the opacity, we can adjust the model to the data. As in Vamvatira-Nakou et al.
(\cite{vamv13}, \cite{vamv15}), the fit is done for data points at $\lambda > $
20 $\mu$m.

The best fit (Fig.~\ref{m1-67_2dust}) was achieved using the following populations
of dust grains and the optical constants of olivines with a 50/50 Fe to Mg abundance
given by Dorschner et al. (\cite{dor95}), extrapolated to a constant refraction index
in the far-ultraviolet (FUV). The first is a population of small grains with radii
from $0.005$ to $0.1$ $\mu$m, which is responsible for the emission at $\lambda < 40 \mu$m.
The second is a population of large grains with radii from $2$ to $10$ $\mu$m, which is
responsible for the infrared SED at $\lambda > 70 \mu$m. Such large grains were also
found in our analysis of the AG Car nebula (Vamvatira-Nakou et al. \cite{vamv15}).

\begin{figure}[t]
\resizebox{\hsize}{!}{\includegraphics*{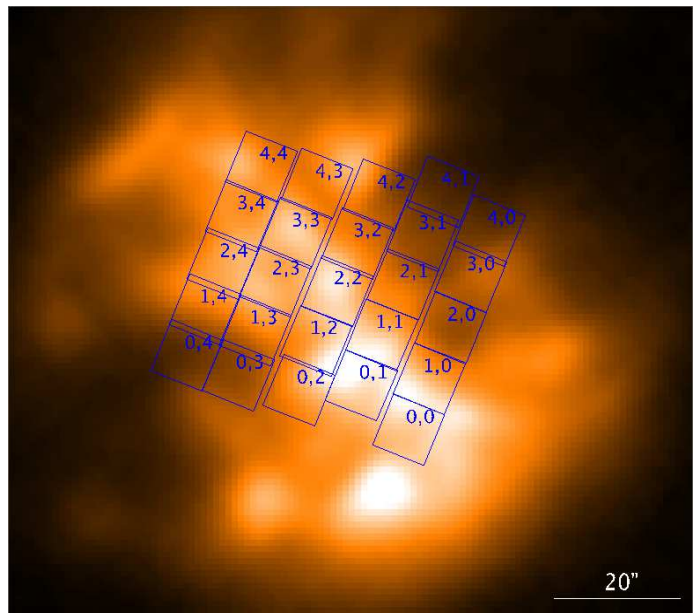}}
\caption{Footprint of the PACS spectral field-of-view on the image 
of the nebula M1-67 at 70\ \mbox{$\mu$m}. This field is composed of
5$\times$5 spaxels, each one of them labeled with a number pair.
North is up and east is to the left.}
\label{m1-67_sp_foot}
\end{figure}

According to the model results, the total mass of dust is $M_{\rm dust}
\sim 0.22$ M$_{\odot}$ ($0.006$ M$_{\odot}$ from the small dust grains
and $0.21$ M$_{\odot}$ from the large dust grains). The uncertainty
of $\sim\ 20\%$  accounts for the dispersion of the mass values
obtained with different models, for example, models that have different grain
composition. The temperature of small grains goes from $65$ K at
$r_{\mathrm{in}}$ to $58$ K at $r_{\mathrm{out}}$, while the temperatures
of large grains goes from $29$ K at $r_{\mathrm{in}}$ to $26$ K at
$r_{\mathrm{out}}$. Cichowolski et al. (\cite{cicho08}) found a dust
temperature of $70$ K for silicates. There is a difference between these
results and ours because we considered two populations of dust grains.

%__________________________________________________________________
\section{Emission line spectrum}
\label{sec:emission line spectrum}

Figure~\ref{m1-67_sp_foot} illustrates the footprint of the PACS spectral
field of view, which is composed of 25 (5$\times$5) spaxels, on the image
of the M1-67 nebula at 70\ \mbox{$\mu$m}. This footprint is not a perfect
square, because one column of spaxels is offset from the others. Each
spaxel corresponds to a different part of the nebula, and we have a spectrum
for each one of them. Nevertheless, a big part of the nebula is outside of
the spectral field-of-view, which is not large enough to cover the entire
nebula.

The integrated spectrum of the nebula over the 25 spaxels is shown in
Fig.~\ref{m1-67_spectra_25sp_NIII57} and Fig.~\ref{m1-67_spectra_25sp},
recalling that the spectroscopic observational mode provides us only with
some preselected areas of the PACS waveband and not a complete coverage. The
forbidden emission lines that have been detected are the following: $[\ion{O}{i}]$
$\lambda\lambda$ 63, 146\ \mbox{$\mu$m}, $[\ion{N}{ii}]$ $\lambda\lambda$ 122,
205\ \mbox{$\mu$m}, $[\ion{C}{ii}]$ $\lambda$ 158\ \mbox{$\mu$m}, $[\ion{N}{iii}]$
$\lambda$ 57\ \mbox{$\mu$m}, and $[\ion{O}{iii}]$ $\lambda$ 88\ \mbox{$\mu$m}.

\subsection{Line flux measurements}

To measure the emission line intensities in each one of the 25 spectra
we performed a Gaussian fit to the line profiles using the Image Reduction
and Analysis Facility (IRAF) package (Tody \cite{tod86}, \cite{tod93}).
These measurements are presented in the Table~\ref{table:4} of
Appendix A. Not all the lines are detected in each spaxel.
The $[\ion{O}{i}]$ 146\ \mbox{$\mu$m} line seems to be the most difficult
to detect. In contrast, the $[\ion{N}{ii}]$ 122\ \mbox{$\mu$m} line,
detected everywhere, has the highest flux among the detected lines in
every spaxel.

\begin{figure}[t]
     \centering
     \includegraphics[width=6.1cm]{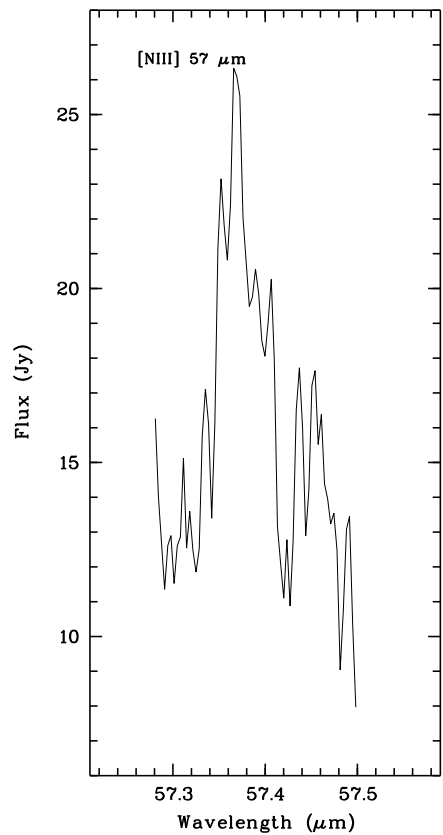}
     \caption{Integrated PACS spectrum of the nebula M1-67 around the WR 124
star over the 25 spaxels. The forbidden emission line $[\ion{N}{iii}]$
57\ \mbox{$\mu$m} is detected.}
      \label{m1-67_spectra_25sp_NIII57}
\end{figure}

\begin{figure*}[!]
  \begin{minipage}[t]{6.1cm}
\centering
    \includegraphics*[width=6.1cm]{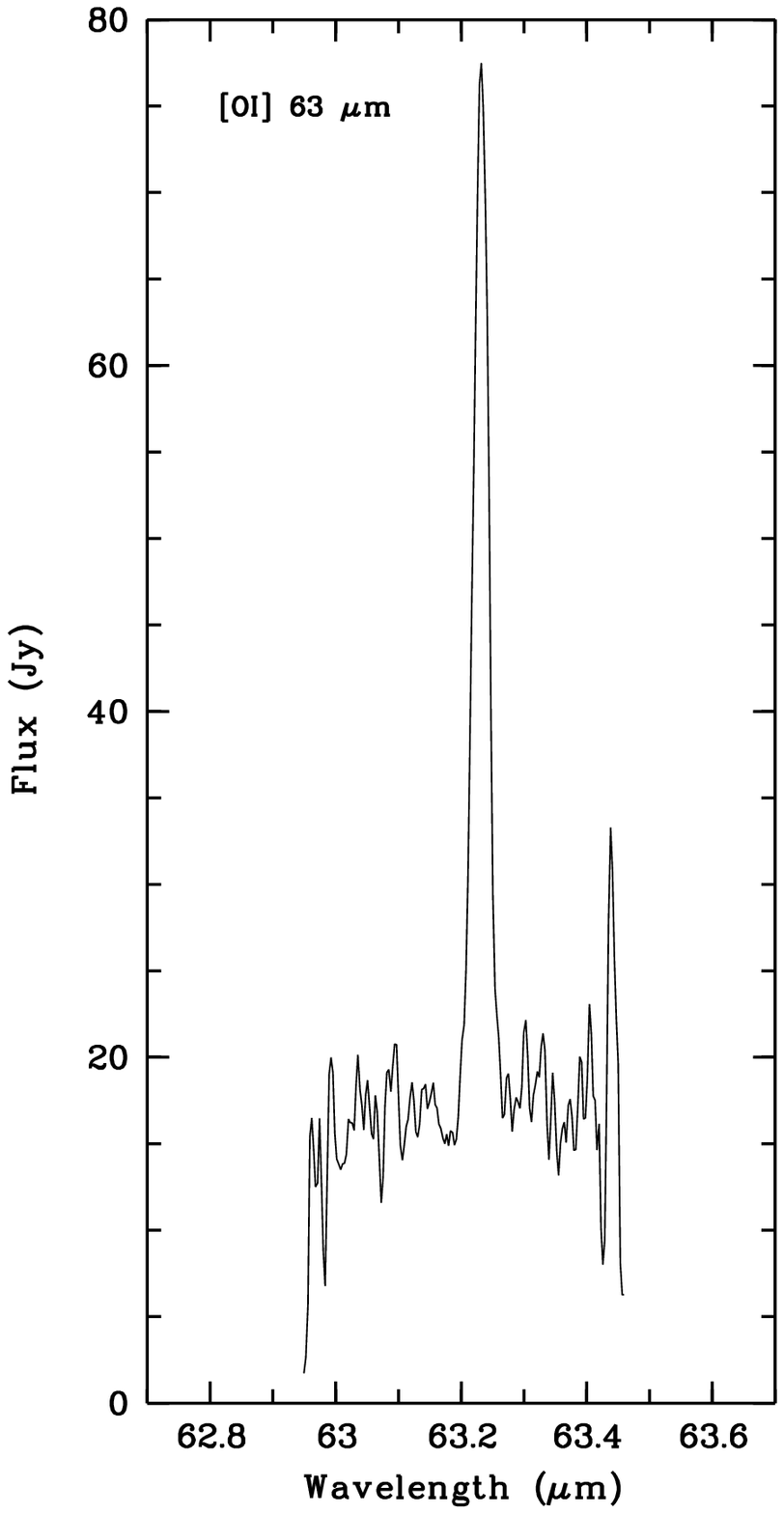}
  \end{minipage}\hfill
\begin{minipage}[t]{6.1cm}
\centering
    \includegraphics*[width=6.1cm]{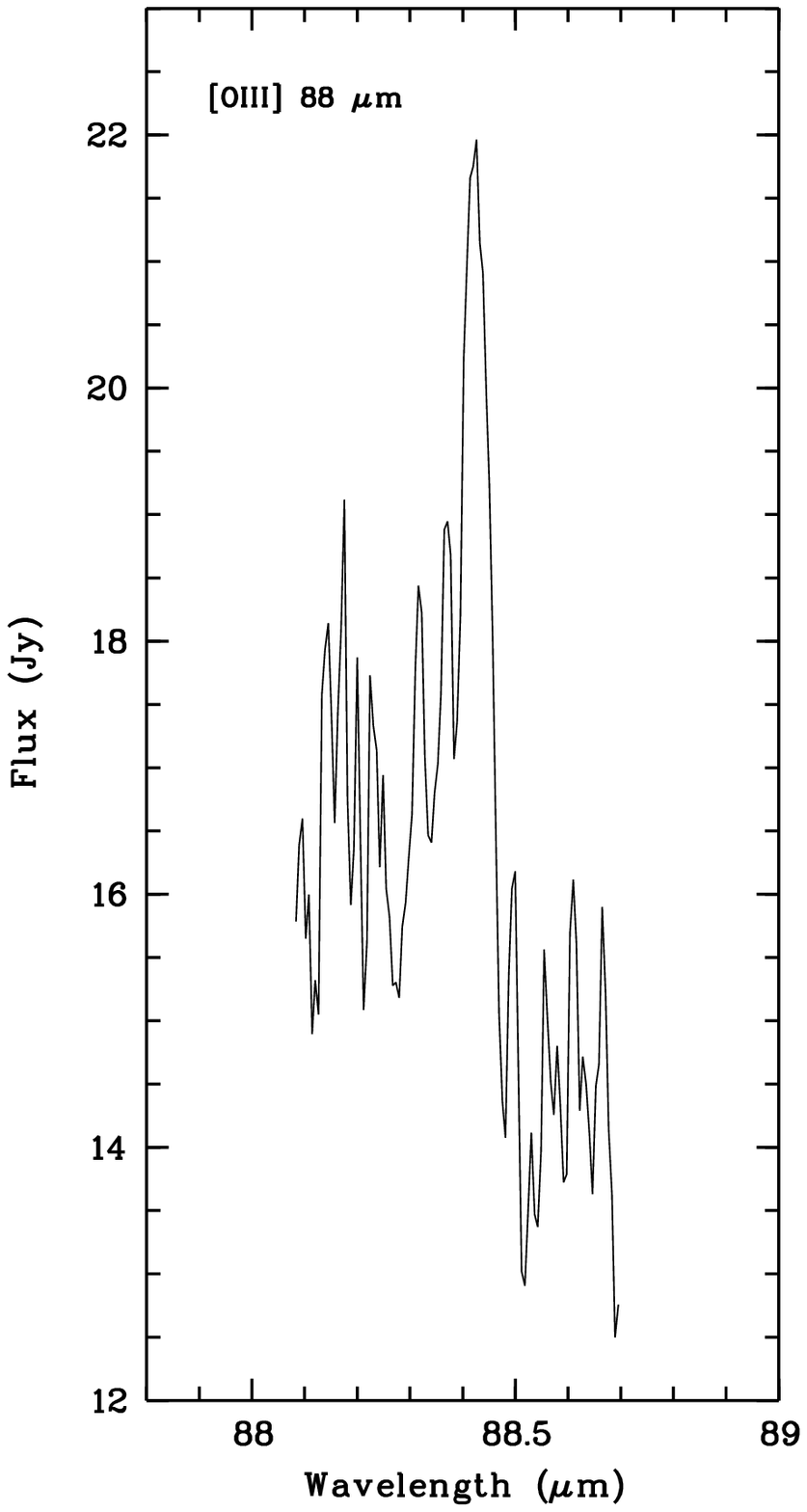}
\end{minipage}\hfill
\begin{minipage}[t]{6.1cm}
\centering
    \includegraphics*[width=6.1cm]{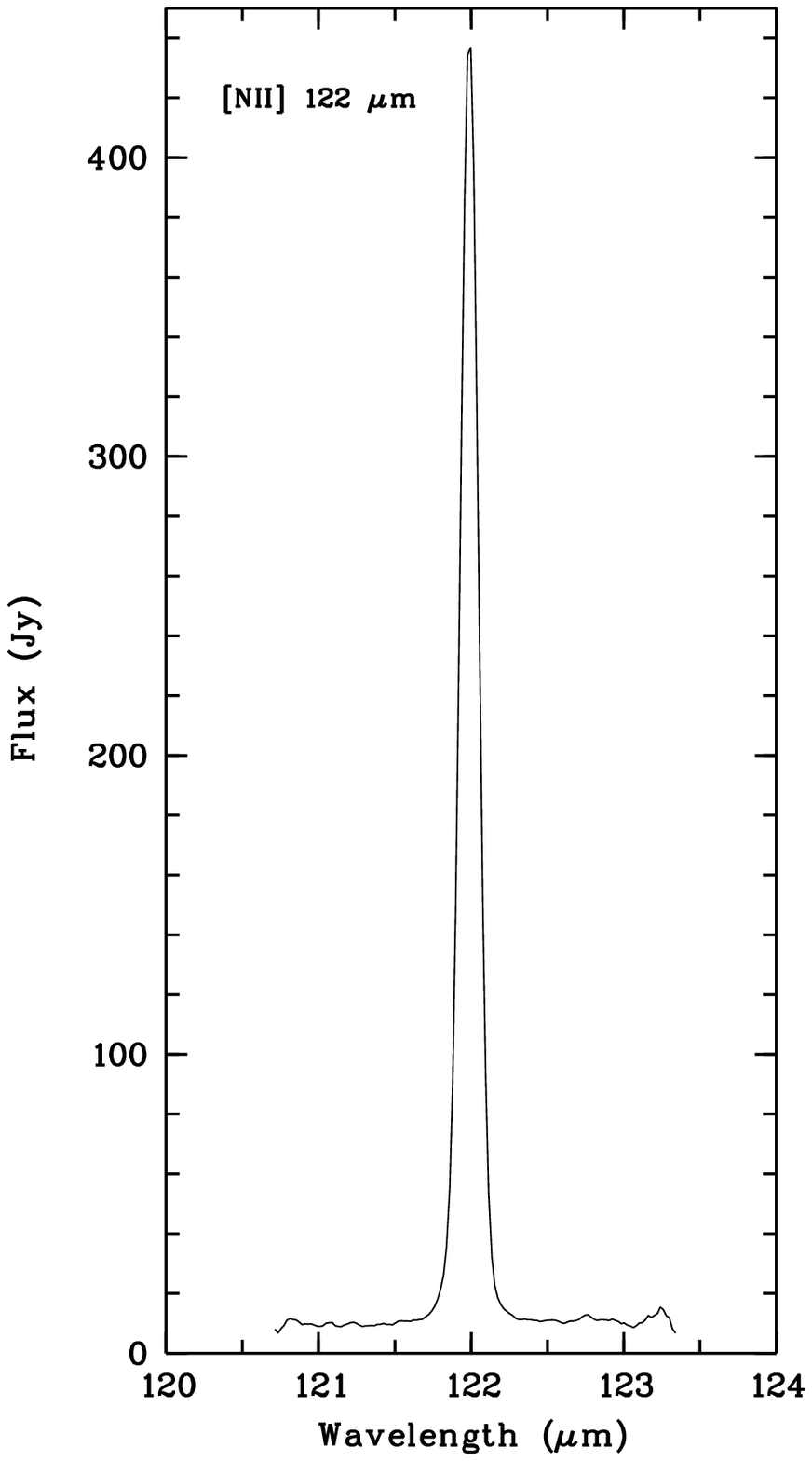}
\end{minipage}\hfill\\
\begin{minipage}[t]{6.1cm}
\centering
    \includegraphics*[width=6.1cm]{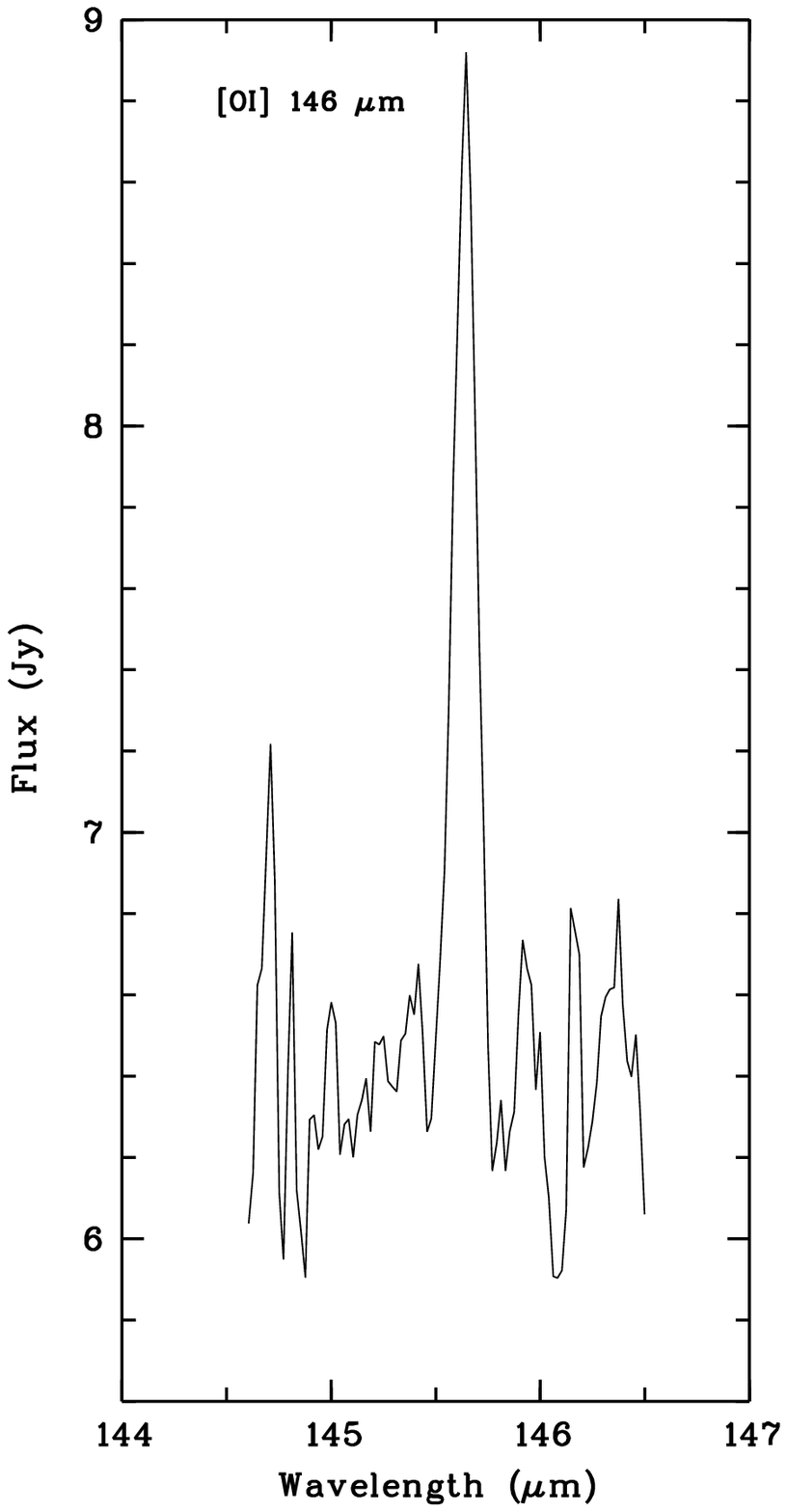}
  \end{minipage}\hfill
\begin{minipage}[t]{6.1cm}
\centering
    \includegraphics*[width=6.1cm]{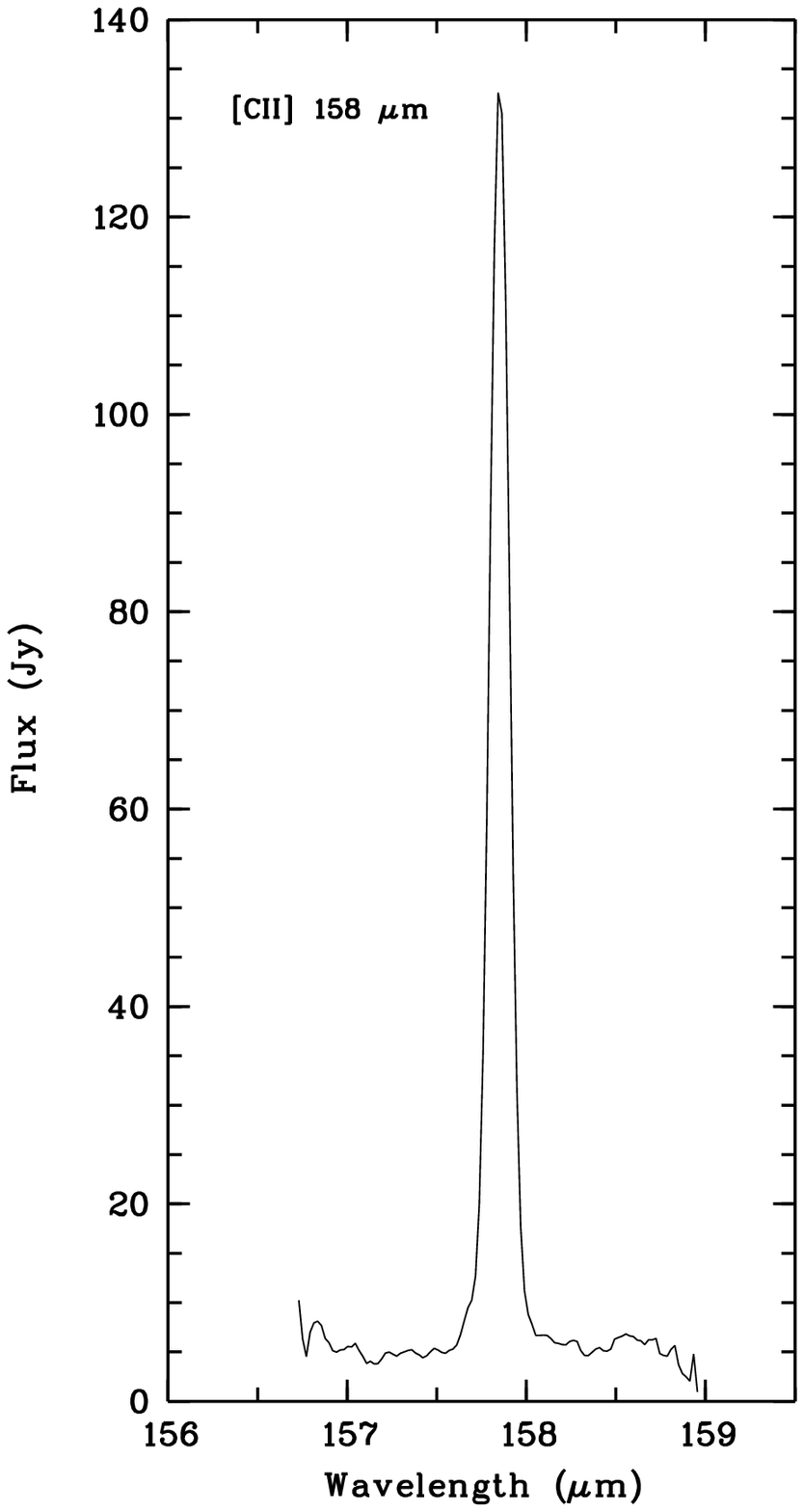}
  \end{minipage}\hfill
\begin{minipage}[t]{6.1cm}
\centering
    \includegraphics*[width=6.1cm]{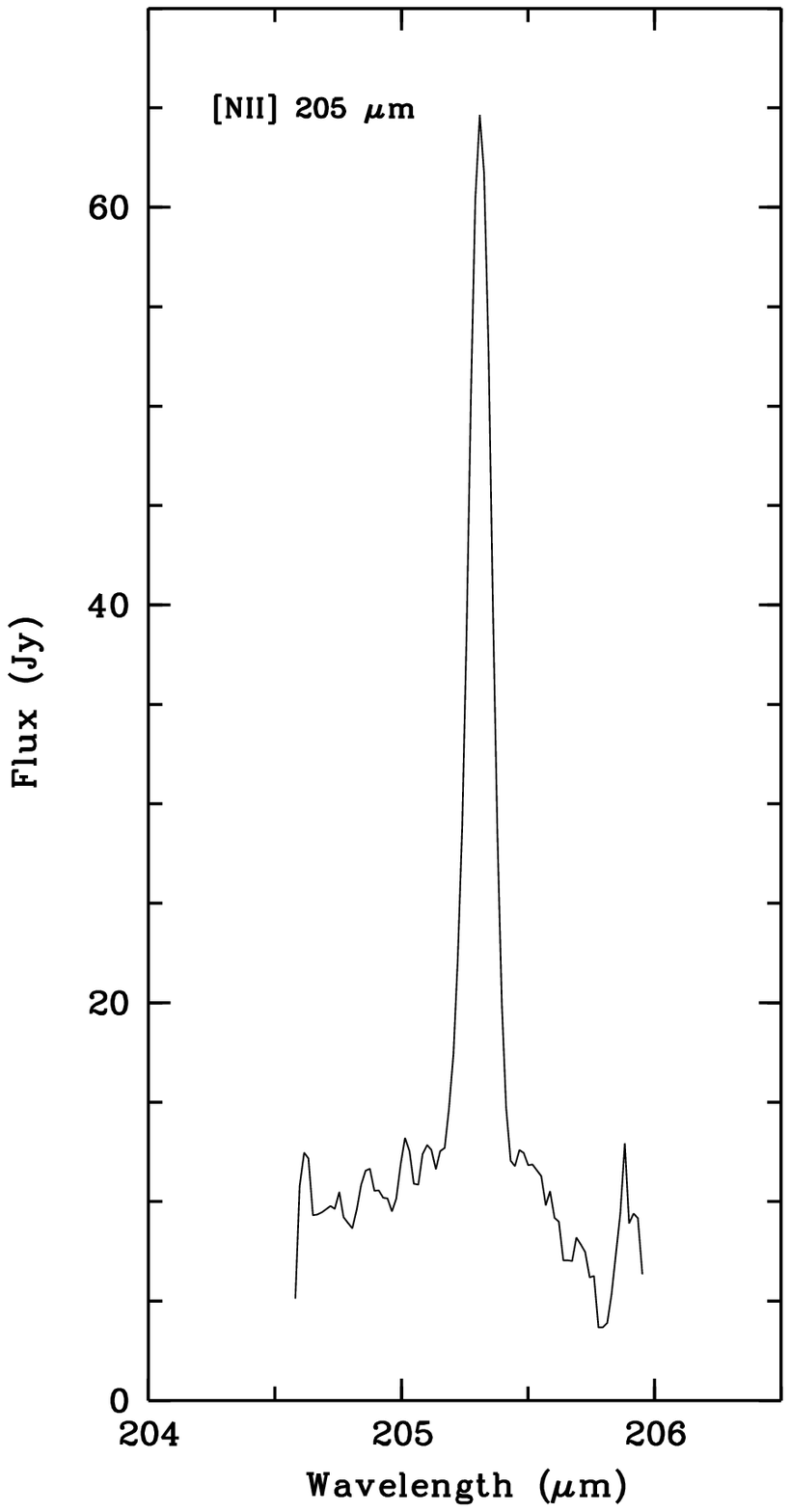}
  \end{minipage}\hfill
    \caption{Integrated PACS spectra of the nebula M1-67 around the WR 124
star over the 25 spaxels. The detected forbidden emission lines are
$[\ion{O}{i}]$ 63\ \mbox{$\mu$m}, $[\ion{O}{iii}]$ 88\
\mbox{$\mu$m}, $[\ion{N}{ii}]$ 122\ \mbox{$\mu$m}, $[\ion{O}{i}]$ 146\
\mbox{$\mu$m}, $[\ion{C}{ii}]$ 158\ \mbox{$\mu$m}, and $[\ion{N}{ii}]$
205\ \mbox{$\mu$m}.}
   \label{m1-67_spectra_25sp}
\end{figure*}

To increase the S/N, we then measured the flux of the emission
lines that are present on the spectrum summed over the 25 spaxels
(Figs.~\ref{m1-67_spectra_25sp_NIII57} and~\ref{m1-67_spectra_25sp}).
We note that the $[\ion{N}{iii}]$ 57\ \mbox{$\mu$m} and $[\ion{O}{iii}]$
88\ \mbox{$\mu$m} lines are only detected in the summed spectrum. The
$[\ion{N}{ii}]$ 205\ \mbox{$\mu$m} line is known to have a problematic
calibration in PACS. As a result, we need to correct its flux before
using it in the following analysis. Using objects from the MESS
collaboration (Groenewegen et al. \cite{groenewegen}) that
were observed with both PACS and SPIRE, we calculated a
correction factor. With the help of the SPIRE/PACS cross
calibration, we then found that the measured $[\ion{N}{ii}]$ 205\ \mbox{$\mu$m}
flux should be multiplied by a correction factor of 4.2, assuming an error
of 25\% for the final corrected $[\ion{N}{ii}]$ 205\ \mbox{$\mu$m} flux.
\footnote{Although there are also SPIRE observations of the
spectrum of the nebula M1-67, taken in the framework of the MESS program,
they cannot be used to have a more precise flux for the line $[\ion{N}{ii}]$
205\ \mbox{$\mu$m}. That is due to the combination of the geometry of the
detector array and the geometry of the nebula and because the
observing mode was a single pointing and not a raster map, in such a way that
any attempt to recover the nebular flux is highly uncertain. As a result,
we decided not to include the SPIRE spectroscopic data in this study.}

The emission line flux measurements from the spectrum summed over the
25 spaxels are given in Table~\ref{table:2}, taking
the correction on the $[\ion{N}{ii}]$ 205\ \mbox{$\mu$m} line flux into account.
There is good agreement between these results and the flux
measurements of the lines $[\ion{N}{ii}]$ $\lambda$ 122\ \mbox{$\mu$m}
and $[\ion{C}{ii}]$ $\lambda$ 158\ \mbox{$\mu$m} that are detected in the 
ISO-LWS spectrum of M1-67 (see Appendix B). Considering that the
circular aperture of this instrument is 80\arcsec so that most of
the bright nebula is inside, we can conclude that
most of the nebular flux is measured when using the PACS spectrum
summed over the 25 spaxels.

\begin{table}[h]
\caption{Line fluxes from the summed spectrum (25 spaxels)
of the nebula M1-67.}
\label{table:2}
\centering
\begin{tabular}{l c c }
\hline\hline
Ion & $\lambda$ & $F\pm\Delta F\ $   \\
    &  ($\mu$m)   & ($10^{-15}$ W~m$^{-2}$)  \\
\hline\hline
    $[\ion{N}{iii}]$ & 57   & 0.29 $\pm$ 0.10  \\
    $[\ion{O}{i}]$   & 63   & 1.34 $\pm$ 0.29  \\
    $[\ion{O}{iii}]$ & 88   & 0.14 $\pm$ 0.04  \\
    $[\ion{N}{ii}]$  & 122  & 13.05 $\pm$ 2.70 \\
    $[\ion{O}{i}]$   & 146  & 0.05 $\pm$ 0.01  \\
    $[\ion{C}{ii}]$  & 158  & 2.13 $\pm$ 0.44  \\
    $[\ion{N}{ii}]$  & 205  & 1.89 $\pm$ 0.47  \tablefootmark{a}\\
\hline
\end{tabular} \\
\tablefoot{
\tablefoottext{a}{Corrected value from PACS/SPIRE cross-calibration}
}
\end{table}

In the following spectral analysis, we only use ratios of the line
fluxes from the spectrum summed over the 25 spaxels to determine nebular
properties, assuming that they represent the whole nebula. We do not use
the line fluxes themselves because they only come from a part of the
nebula that corresponds to the area covered by the 25 spaxels.

\subsection{Photoionization region characteristics}

The following four detected emission lines: $[\ion{N}{ii}]$ 122,
205\ \mbox{$\mu$m}, $[\ion{N}{iii}]$ 67\ \mbox{$\mu$m}, and
$[\ion{O}{iii}]$ 88\ \mbox{$\mu$m} are
associated to the photoionization region (i.e., \ion{H}{ii} region)
of the M1-67 nebula. The other forbidden emission lines, detected in
the nebular spectrum, originate in a region of transition between
ionized and neutral hydrogen. They may indicate the presence of a
photodissociation region (PDR), which we analyze and discuss in
the next section.

\subsubsection{H$\alpha$ flux}

To estimate the H$\alpha$+$[\ion{N}{ii}]$ flux from the M1-67 nebula
we integrated the surface brightness over the whole nebula. We
corrected for the contamination by field stars and the background
and we extrapolated the emission from the occulted central part
using the mean surface brightness. We measured the continuum flux
from the reflection nebula in the adjacent filter, accounting
for the difference in filter transmissions. Since WR 124 is a strong
emission-line star, the reflected stellar H$\alpha$ flux must also be
subtracted. The final contamination due to the reflection nebula was
estimated to be 13\%, considering the H$\alpha$ equivalent widths
measured by Hamann et al. (\cite{ham93}) for WR 124 in 1992
(i.e., accounting for $\sim$ 3 years of time delay). We subtracted
the contribution of the strong $[\ion{N}{ii}]$ lines using the
$[\ion{N}{ii}]$ /H$\alpha$ ratios from available spectroscopic
data and from the transmission curve of the H$\alpha$+$[\ion{N}{ii}]$
filter. With the help of the three PN and the three
spectrophotometric standard stars observed in the same filter, we
did the conversion to absolute flux. The conversion factors derived
from all these six objects are in excellent internal agreement.

The H$\alpha$ flux was measured to be $F(\mathrm{H}\alpha$) = 2.4
$\times$ 10$^{-11}$ ergs~cm$^{-2}$~s$^{-1}$ uncorrected for reddening
and with an uncertainty of $\sim$ 20\%. Adopting E(B$-$V ) = 0.93 $\pm$
0.10 (Esteban et al. \cite{est91}), we derived $F_{0}(\mathrm{H}\alpha$)
= (2.0 $\pm$ 0.6) $\times$ 10$^{-10}$ ergs~cm$^{-2}$~s$^{-1}$ for the
M1-67 nebula. Within the uncertainties, this flux agrees with
the flux measured by Grosdidier et al. (\cite{gro98}): $F_{0}(\mathrm{H}
\alpha$) = 2.3 $\times$ 10$^{-10}$ ergs~cm$^{-2}$~s$^{-1}$ using
the same reddening. The flux density from the reflection nebula is
$F_{\lambda}$ = 2.9 $\times$ 10$^{-14}$ ergs~cm$^{-2}$~s$^{-1}$
~\AA$^{-1}$ at 6650~\AA\ (the central wavelength of the continuum
filter).

The Esteban et al. (\cite{est91}) reddening estimate was based
on long-slit Balmer line spectral ratios. The wavelength baseline used is
small, and considering how dusty this nebula is, there may well be significant
reddening variations across the nebula. Given that there is a good S/N
8.64 GHz radio flux of $205 \pm 10$ mJy from 
Cichowolski et al. (\cite{cicho08}), we can get a more accurate reddening-free
H$\alpha$ flux estimate from that flux using the radio-Balmer method of Milne
and Aller (\cite{milne75}). Using the formulae given in Vamvatira-Nakou et al.
(\cite{vamv13}) that are based on Osterbrock \& Ferland (\cite{oster06})
and Draine (\cite{draine11}), we can write
       \begin{equation} \label{eq:radio_to_Halpha_m1-67}
       S_{\nu}/F(\mathrm{H}\alpha) 
                = 9.9\times10^{11}(1 + y_{+})\left(\frac{4.9}{\nu}
                     \right)^{0.1}T_4^{(0.59+0.03\mathrm{ln}T_4)}
       \end{equation} 
where the radio flux, $S_{\nu}$, is in mJy; the H$\alpha$ flux is in ergs~cm$^{-2}$~s$^{-1}$,
$y_{+} = n_{\mathrm{He^{+}}}/n_{\mathrm{H^{+}}}$, with
$n_{\mathrm{H}^{+}}=n_{\mathrm{p}}$, $n_{\mathrm{He}^{+}}$ being the number
densities of the ionized hydrogen and ionized helium, respectively; the radio
frequency, $\nu$, is in GHz; and $T_4=T_e/(10^4\ \mathrm{K})$.
Assuming an electron temperature, $T_\mathrm{e}$, equal to $7000$ K (mean value
of the estimate of Barker \cite{bark78} and Esteban et al. \cite{est91}) that
we adopt for the following analysis as Fern\'{a}ndez-Mart\'{i}n et al. (\cite{fern13})
with an uncertainty of 20\% that is constant throughout the nebula, we have
$T_4 = 0.7$. The value of $y_{+} = 0.013$ is taken from Esteban et al.
(\cite{est91}). This  gives an unreddened H$\alpha$ flux of $F_{0} = 2.7\times 10^{-10}$
ergs~cm$^{-2}$~s$^{-1}$ with an uncertainty of about 30\%. This is the
value we use for the following analysis.

\subsubsection{Electron density}

The $[\ion{S}{ii}]$ 6717/6731 ratio is an electron density diagnostic
in the optical. Several electron density estimates were made based
on this ratio. Solf and Carsenty (\cite{solf82}) calculated an electron
density of $1000 \pm 300\ \mathrm{cm}^{-3}$. Later on, Esteban et al.
(\cite{est91}) calculated  the electron density in different parts of
the nebula (four different slit positions). Their results for the central
parts agree with the value of Solf and Carsenty (\cite{solf82}), while
in the outer parts the electron density is about $200$ cm$^{-3}$. The
results of Sirianni et al. (\cite{sir98}), who presented a detailed
electron density distribution versus the distance from the star, confirmed
the previous two studies. Fern\'{a}ndez-Mart\'{i}n et al. (\cite{fern13})
calculated a density range from $\sim$ $1500$ cm$^{-3}$, near the star, to
$\sim$ $650$ cm$^{-3}$ toward the nebular edge, results that are consistent
with their electron density maps and the previous studies.

Grosdidier et al. (\cite{gro98}) estimated the electron density to
be $825$ $\pm 115\ \mathrm{cm}^{-3}$ based on the analysis of H$\alpha$
observations of the nebula and adopting a distance of $4.5$ kpc.
Cichowolski et al. (\cite{cicho08}) found a value of $630-360$ cm$^{-3}$
based on radio continuum data at 3.6 cm and adopting a distance of $5$ kpc.
Considering the uncertainties, these estimations agree with the previously
mentioned ones that were based on the $[\ion{S}{ii}]$ 6717/6731 ratio.

The $[\ion{N}{ii}]$ 122/205\ \mbox{$\mu$m} ratio is a diagnostic
for the electron density of the nebula in the infrared waveband.
Using the package \textit{nebular} of the IRAF/STSDAS environment
(Shaw \& Dufour \cite{shaw}), and considering the adopted electron
temperature and the values given in Table~\ref{table:2}, the electron
density of the nebula M1-67 is estimated to be $600 \pm 180\
\mathrm{cm}^{-3}$. The uncertainty was estimated and accounts
for the measurement uncertainties and for the dispersion of values
calculated using different electron temperatures within errors.
Given the uncertainties, this estimate of the
average nebular electron density agrees with the previous
studies. Furthermore, when a nebula has a spatially inhomogeneous
electron density, the use of different line ratios as density
diagnostics lead to different results because of the difference in
critical density between the lines used for the density calculation
(Rubin \cite{rub89}, Liu et al. \cite{liu01}). This has been
observed in PN (Liu et al. \cite{liu01}, Tsamis
et al. \cite{tsam03}) and in a LBV nebula (Vamvatira-Nakou et al.
\cite{vamv15}).

Our estimate of the electron density based on infrared data will be
used in the following calculations, because the best determination of
the electron density is done when it is similar to the critical
density of the lines used for the calculation (Rubin et al. \cite{rub94}).
Otherwise, the further calculation of ionic abundances will not give
correct results (Rubin \cite{rub89}, Liu et al. \cite{liu01}).

It should be mentioned that the electron density was also calculated
in each spaxel separately, in an effort to examine whether there are any spatial
differences. Given the large uncertainties of the flux measurements,
no significant spatial variation of the electron density was detected.

\subsubsection{Ionizing flux}

Using the estimated H$\alpha$ flux, the rate of emission of hydrogen-ionizing
photons, $Q_{0}$, in photons $\mathrm{s}^{-1}$, and the radius of the
Str\"omgren sphere, $R_{S}$, in pc, can be estimated with the help of the
following equations (Vamvatira-Nakou et al. \cite{vamv13})
       \begin{equation} \label{eq:Q_H alpha_m1-67}
        Q_{0(\mathrm{H\alpha})}=8.59\times10^{55}
                  T_4^{(0.126+0.01\mathrm{ln}T_4)}D^2F_0(\mathrm{H}_{\alpha}) \; ,
       \end{equation}
       \begin{equation} \label{eq:stromgren radius_m1-67}
        R_{\mathrm{S}}=3.17\left(\frac{x_e}{\epsilon}\right)^{1/3}
                      \left(\frac{n_\mathrm{e}}{100}\right)^{-2/3}
                       T_4^{(0.272+0.007\mathrm{ln}T_4)}\left
                       (\frac{Q_0}{10^{49}}\right)^{1/3}\; ,
       \end{equation}
 where $x_{\mathrm{e}}=n_{\mathrm{e}}/n_{\mathrm{p}}\simeq1+n_{\mathrm{He^{+}}}/n_{\mathrm{H^{+}}}=1+y_{+}$, 
assuming
that the number density of the doubly ionized helium is $n_{\mathrm{He^{++}}}=0$,
$\epsilon$ is the filling factor, D
the distance of the nebula in kpc, and $F_0(H\alpha)$ the H$\alpha$ flux
in ergs~cm$^{-2}$~s$^{-1}$.

Assuming $\epsilon = 0.05$, which is the value used by Grosdidier
et al. (\cite{gro98}) and using the abundance ratio
$n_{\mathrm{He^{+}}}/n_{\mathrm{H^{+}}} = 0.013$ from Esteban et al. (\cite{est91})
and the adopted value for  $T_\mathrm{e}$,
we calculated a rate of emission of hydrogen-ionizing photons of about
$2.7 \times 10^{47}$ photons $\mathrm{s}^{-1}$ and a Str\"omgren radius
of $0.8$ pc. For a higher filling factor of $0.15$ (Cichowolski et al. \cite{cicho08}),
the Str\"omgren radius is $1.1$ pc. Considering the uncertainties of the nebular
parameters, our estimate for the Str\"omgren radius agrees with the
value of $1.3$ pc based on radio measurements  (Cichowolski et al. \cite{cicho08}),
given that they used a higher distance ($5$ kpc).

The ionized gas nebula extends up to about $0.9$ pc from the central star
(Sect.~\ref{sec:morphology of the nebula}). This value is comparable, considering
the uncertainties, to the estimated Str\"omgren radius that is the radius of
an ionization bounded nebula by definition. Consequently, the nebula M1-67
around WR 124 may be ionization bounded.

\subsubsection{Abundance ratio N/O} 

\begin{figure*}[t]
\sidecaption
  \includegraphics[width=12cm]{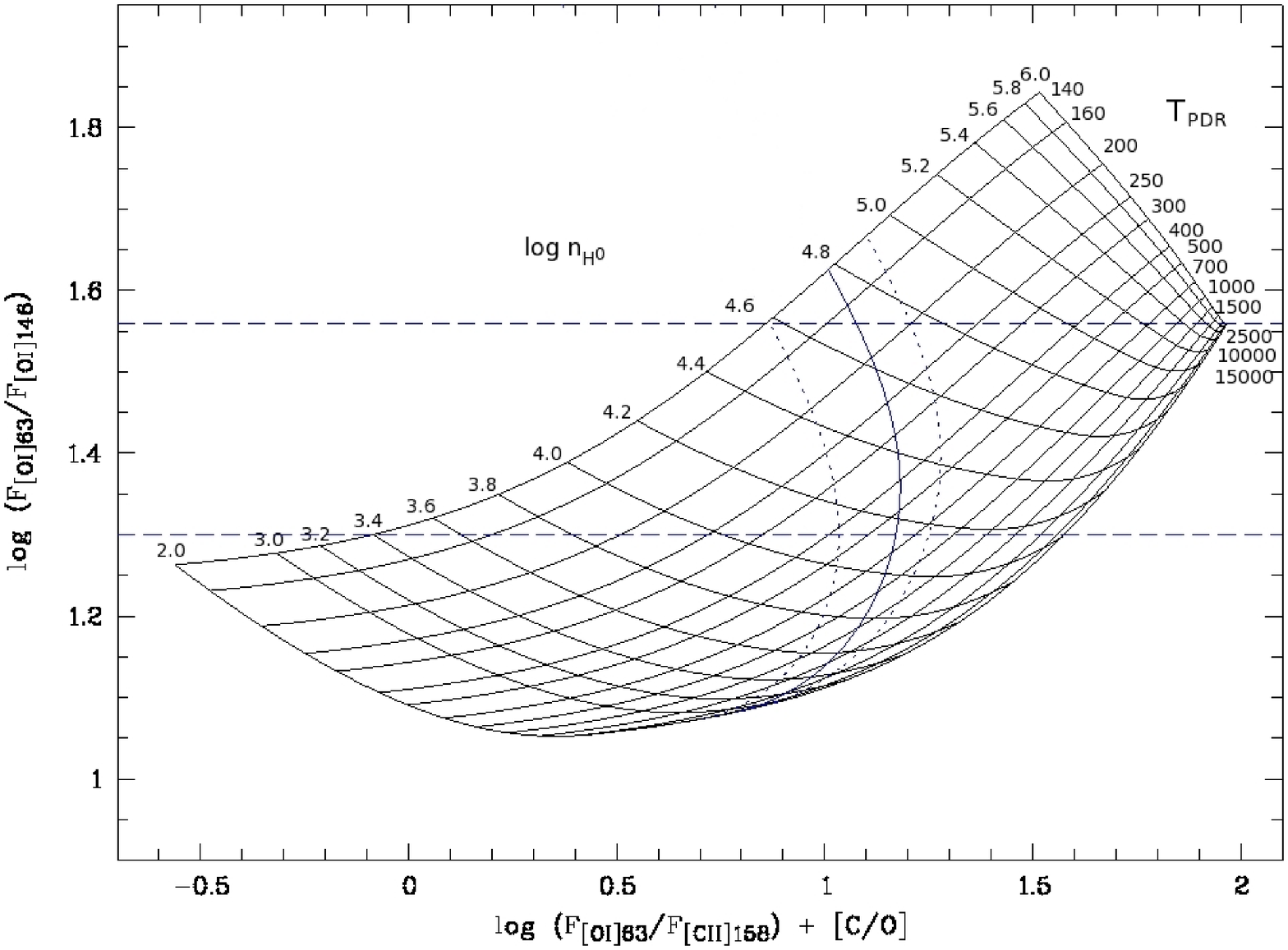}
     \caption{Temperature-density PDR diagnostic diagram. The grid
of flux ratios $F_{[O{\textsc {i}}]63}/ F_{[O{\textsc {i}}]146}$ versus
$F_{[O{\textsc {i}}]63}/F_{[C{\textsc {ii}}]158}^{\mathrm{PDR}}$ was
calculated by solving the level population equations for a range of
temperatures and densities. $F_{[O{\textsc {i}}]63}/F_{[C{\textsc {ii}}]158}
^{\mathrm{PDR}}$ is normalized to the solar abundance (C/O)$_{\odot} = 0.5$
so that [C/O] $\equiv$ log(C/O) - log(C/O)$_{\odot}$. The solid line corresponds
to the pressure equilibrium between the H{\sc ii} region and the PDR, the two
dotted lines on each side accounting for the errors. The
horizontal dashed lines correspond to the observational
log$(F_{[O{\textsc {i}}]63}/F_{[O{\textsc {i}}]146})$ ratio with its error.}
     \label{lineratio_m1-67}
\end{figure*}

Using the emission lines $[\ion{N}{iii}]$ 57\ \mbox{$\mu$m} and $[\ion{O}{iii}]$
88\ \mbox{$\mu$m}, the N/O abundance number ratio can be estimated by
the equation
      \begin{equation}\label{eq:NO ratio abundance m1-67}
          \frac{\mathrm{N}}{\mathrm{O}}=\frac{\langle \mathrm{N}^{++}\rangle}
                        {\langle \mathrm{O}^{++}\rangle}
                       =\frac{F_{[N{\textsc {iii}}]57}/\varepsilon_{[N{\textsc {iii}}]57}}
                             {F_{[O{\textsc {iii}}]88}/\varepsilon_{[O{\textsc {iii}}]88}} \,,
      \end{equation}
where $F$ is the flux and $\varepsilon$ is the volume emissivity  of a given
line. The emissivities were calculated with the package \textit{\emph{"nebular"}},
using the adopted value for the electron temperature and the previously calculated
electron density. From the measured line fluxes and their
uncertainties (Table~\ref{table:2}), the
N/O abundance ratio is $1.0 \pm 0.5$, a value that is much higher than the solar
one, $0.14$ (Grevesse et al. \cite{grev10}). Given the uncertainties, it is in
agreement with the value of Esteban et al. (\cite{est91}) and similar
to the values derived for other nebulae around WR stars (Smith \cite{smith97}).

\subsubsection{Mass of the ionized gas}

Based on the H$\alpha$ emission we can estimate the ionized gas mass. Assuming
a spherical nebula, the ionized mass in solar masses is given by (Vamvatira-Nakou
et al. \cite{vamv13})
       \begin{equation} \label{ionized mass H alpha_sphere m1-67}
              M_{i(\mathrm{H}\alpha)}=57.9\frac{1+4y_{+}}{\sqrt{1+y_{+}}}T_4^{(0.471+0.015\mathrm{ln}T_4)}\epsilon^{1/2}
                   \theta^{3/2}D^{5/2}F^{1/2}_0(\mathrm{H}\alpha),
       \end{equation}
where $\theta$ is the angular radius of the nebula ($R=\theta D$) in arcsec and
$y_{+}=n_{\mathrm{He^{+}}}/n_{\mathrm{H^{+}}}$.

Using the previous assumptions for the filling factor and the ${\mathrm{He^{+}}}/{\mathrm{H^{+}}}$
abundance ratio, we estimated the mass of the ionized gas to be $1.5 \pm\ 0.8\ \mathrm{M}_{\odot}$.
For the higher filling factor, $\epsilon = 0.15$, the ionized gas mass is
$2.7 \pm\ 1.4 $ M$_{\odot}$.

Previous studies gave various results for the mass of ionized gas in the nebula that
do not always agree with each other because different assumptions are made
and different parameters are used. Cohen and Barlow (\cite{coh75}) assumed a filling
factor of unity to obtain their ionized mass of $9$ M$_{\odot}$.
Solf and Carsenty (\cite{solf82}) estimated a mass of about $0.8$ M$_{\odot}$, which
as they reported, agrees with the value of Johnson (\cite{john80}) and
Chu and Treffers (\cite{chu81}). Later on, Grosdidier et al. (\cite{gro98}) estimated
that the total mass of the ionized shell could be as low as about $1.33$ M$_{\odot}$.
More recently, Cichowolski et al. (\cite{cicho08}) calculated a mass of $5-9$ M$_{\odot}$,
depending on the filling factor used ($0.05-0.15$).

If we consider the geometry of the nebula as described in Sect.~\ref{sec:dust
continuum emission}, the shell has an inner radius of $\theta_{\mathrm{in}}$ = 40\arcsec
and an outer radius $\theta_{\mathrm{out}}$ = 60\arcsec. The mass of the ionized shell
nebula is then given by
       \begin{equation} \label{ionized mass H alpha_shell m1-67}
               M_{i}^{\mathrm{shell}}=(\theta_{\mathrm{out}}^3-\theta_{\mathrm{in}}^3)^{1/2}\theta_{\mathrm{out}}^{-3/2}M_{i}^{\mathrm{sphere}} .
       \end{equation}
The mass of the ionized shell nebula is thus $1.3 \pm 0.7 \ \mathrm{M}_{\odot}$ 
for $\epsilon = 0.05$ and $2.3 \pm 1.2 \ \mathrm{M}_{\odot}$
for $\epsilon = 0.15$. The quoted errors were estimated by propagating
the uncertainties of the used quantities.

\subsection{Photodissociation region characteristics}

The detection of the other three emission lines, $[\ion{O}{i}]$ 63, 146\ \mbox{$\mu$m},
and $[\ion{C}{ii}]$ 158\ \mbox{$\mu$m}, may indicate the presence of a PDR in
the nebula, because these fine structure lines are among the most important coolants
in PDRs (Hollenbach \& Tielens \cite{holl97}). Nevertheless, a shock, resulting from
the fast stellar wind that interacts with the slow expanding remnant formed in a
previous stage of the stellar evolution, could also be the cause of the observed
$[\ion{O}{i}]$ and $[\ion{C}{ii}]$ emission. In either case, the ratio $[\ion{O}{i}]$
63\ \mbox{$\mu$m} / $[\ion{C}{ii}]$ 158\ \mbox{$\mu$m} can be used to exclude one of
these two possible scenarios. This ratio is $\gtrsim$ 10 in shocks (Hollenbach and
McKee \cite{holl89}, Castro-Carrizo et al. \cite{castr01}), which is not the case
for the nebula M1-67, where this ratio is equal to $0.6$. Consequently, we can
conclude that a PDR is probably responsible for the $[\ion{O}{i}]$ and $[\ion{C}{ii}]$
emission detected in the nebula M1-67.

In the past, PDRs were detected in some nebulae that surround LBV stars.
First, Umana et al. (\cite{uman09}) discovered a PDR in the nebula around the
LBV HR Car. Later on, Umana et al. (\cite{uman10}) observed a PDR in the nebula
associated with the LBV candidate HD 168625. The analysis of \textit{Herschel}
data revealed the presence of a PDR in the nebula around the LBV WRAY 15-751
(Vamvatira-Nakou et al. \cite{vamv13}) and around the LBV AG Car (Vamvatira-Nakou
et al. \cite{vamv15}).

It should be stressed here that the data were checked for possible background
contamination and that there is no significant contribution from the background
to none of the observed emission lines. The measured line fluxes come entirely
from the nebula M1-67.

The C/O abundance ratio can be estimated based on the PDR line fluxes and
following the method described in Vamvatira-Nakou et al. (\cite{vamv13})
that was used to disentangle the contribution of the PDR and the \ion{H}{ii}
region to the flux of $[\ion{C}{ii}]$ 158\ \mbox{$\mu$m}. In the ionized gas region,
the ratio of fractional ionization is given by
\begin{equation}\label{eq:fr ionization CN m1-67}
      \frac{\langle \mathrm{C}^{+}\rangle}{\langle \mathrm{N}^{+}\rangle}=
        \frac{F_{[C{\textsc {ii}}]158}^{H{\textsc {ii}}}/\varepsilon_{[C{\textsc {ii}}]158}}
                {F_{[N{\textsc {ii}}]122}/\varepsilon_{[N{\textsc {ii}}]122}} \; ,
\end{equation}
where we define $F_{[C{\textsc {ii}}]158}^{H{\textsc {ii}}} =\alpha F_{[C{\textsc {ii}}]158}$,
with $F_{[C{\textsc {ii}}]158}$ being the total flux of the [C{\sc ii}]
158\ \mbox{$\mu$m} line from Table~\ref{table:2} and $\alpha$ a factor
to be determined. Assuming that $\langle \mathrm{C}^{+}\rangle/\langle
\mathrm{N}^{+}\rangle=$ C/N and calculating the emissivities using the
package \textit{nebular}, we end up with the following equation
\begin{equation} \label{eq:cnratio m1-67}
    \frac{F_{[C{\textsc {ii}}]158}^{H{\textsc {ii}}}}{F_{[N{\textsc {ii}}]122}}=
                     (0.34 \pm 0.02) \ \frac{\mathrm{C}}{\mathrm{N}} \; .
\end{equation} 
Since N/O has been estimated to be $\sim 1$, we have
\begin{equation} \label{eq:cnratio2 m1-67}
      \log\alpha = \log\frac{\mathrm{C}}{\mathrm{O}} + 0.32 \; ,
 \end{equation}
using the observed ratio $F_{[C{\textsc {ii}}]158}/F_{[N{\textsc {ii}}]122}$
= $0.163 \pm 0.048$.

The theoretical $F_{[O{\textsc {i}}]63}/ F_{[O{\textsc {i}}]146}$ ratio against
the $F_{[O{\textsc {i}}]63}/F_{[C{\textsc {ii}}]158}^{\mathrm{PDR}}$ ratio normalized
to the solar (C/O)$_{\odot} = 0.5$ abundance ratio is plotted in Fig.~\ref{lineratio_m1-67},
following the study by Liu et al. (\cite{liu01}),
so as to derive the C/O abundance ratio, as well as the temperature, $T_\mathrm{PDR}$,
and the density, $n_{\mathrm{H}^0}$, of the PDR (for details see Vamvatira-Nakou
et al. \cite{vamv13})\footnote{The difference between the diagnostic diagram of Liu et
al. (\cite{liu01}) and ours is due to the use of updated collision coefficients.
We also consider a wider range of density and temperature values.}.
Assuming that there is pressure equilibrium between the ionized gas region and the PDR,
we have the following relation (Tielens \cite{tie05})
\begin{equation} \label{eq:equilibrium m1-67}
    n_{\mathrm{H}^0}kT_\mathrm{PDR}\simeq 2n_\mathrm{e}kT_\mathrm{e}
           =(8.4\pm3.0)\times10^6\ \mathrm{cm}^{-3}\mathrm{K} \; ,
\end{equation}
which is used to define a locus of possible values in the diagram of Fig.~\ref{lineratio_m1-67}.

From Fig.~\ref{lineratio_m1-67} we derive log$(F_{[O{\textsc {i}}]63}$/ $F_
{[C{\textsc {ii}}]158}^{\mathrm{PDR}})$ + [C/O] = $1.1$ where by definition [C/O]
$\equiv$ log(C/O) - log(C/O)$_{\odot}$, using the constraints from
Eq.~\ref{eq:equilibrium m1-67} and the observed ratio $F_{[O{\textsc
{i}}]63}/F_{[O{\textsc {i}}]146} = 26.8 \pm 7.9$. Finally, also using
Eq.~\ref{eq:cnratio2 m1-67}, we calculate $\alpha = 0.95 \pm 0.03$ and C/O =
$0.46\pm 0.27$, which is the solar abundance ratio considering the errors. From the
calculated C/O and N/O abundance ratios and considering the N/H abundance ratio
of Esteban et al. (\cite{est91}), the C/H abundance ratio is then
$(1.3 \pm 0.9) \times 10^{-4}$. The contribution
of the \ion{H}{ii} region to the $[\ion{C}{ii}]$ 158\ \mbox{$\mu$m} line flux is then
$F_{[C{\textsc {ii}}]158}^ {H{\textsc {ii}}} =(2.02\pm0.42)\times 10^{-15}$ W~m$^{-2}$,
while the contribution of the PDR is $F_{[C{\textsc {ii}}]158}^{\mathrm{PDR}} = 
(0.11\pm0.07)\times 10^{-15}$ W~m$^{-2}$, recalling that these are the fluxes of
the integration on the 25 spaxels and not the fluxes from the whole nebula.

The values of the density and the temperature of the PDR
are also provided by the diagram in Fig.~\ref{lineratio_m1-67}. Using the
observed $F_{[O{\textsc {i}}]63}/F_{[O{\textsc {i}}]146}$ ratio we find $\log n_
{\mathrm{H}^0} = 4.43 \pm 0.33$ and $T_\mathrm{PDR} \simeq 470 \pm 330$ K.

The total mass of hydrogen in the PDR, $M_{\mathrm{H}}$, can be estimated from
the $[\ion{C}{ii}]$ 158\ \mbox{$\mu$m} line flux, using the equation given in
Vamvatira-Nakou et al. (\cite{vamv13}) that is
based on Tielens (\cite{tie05}). For the above PDR density,
temperature, distance and C/H abundance, the neutral hydrogen mass is estimated to
be $M_{\mathrm{H}}= 0.05 \pm 0.02\ \mathrm{M}_{\odot}$, which is a lower limit, since
the 25 spaxels do not cover the whole nebula.

Since we naturally expect the PDR to be located just outside the $\ion{H}{ii}$
region, we checked for spatial variation of the  $[\ion{O}{i}]$ 63\ \mbox{$\mu$m} /
$[\ion{C}{ii}]$ 158\ \mbox{$\mu$m} line ratio. In three spaxels that correspond to
nebular clumps this ratio is twice bigger than the average ratio. But the $[\ion{O}{i}]$
63\ \mbox{$\mu$m} / $[\ion{O}{i}]$ 146\ \mbox{$\mu$m} is constant, given the large
uncertainties. Moreover, when the method of disentangling the contribution of the PDR
and the $\ion{H}{ii}$ to the flux of $[\ion{C}{ii}]$ 158\ \mbox{$\mu$m} is performed on
the brightest central spaxel (2,2), i.e. where the $[\ion{O}{i}]$ 63\ \mbox{$\mu$m} /
$[\ion{C}{ii}]$ 158\ \mbox{$\mu$m} line ratio has the highest value, the result shows
that the contribution of the PDR is not higher than the average over the 25 spaxels.
Consequently, we conclude that no variation is detected, which is not surprising since
the $[\ion{C}{ii}]$ line is dominated by the $\ion{H}{ii}$ region contribution and since
the outskirts of the nebula are not in the field of view. On the other hand we cannot
exclude that the PDR emission comes from dense clumps inside the $\ion{H}{ii}$ region
as in the case of the planetary nebula NGC 650 (van Hoof et al. \cite{vanH13}).

%______________________________________________________________
\section{Discussion}
\label{sec:discussion}

The parameters of WR 124 and its nebula M1-67
are summarized in Table~\ref{table:3}. The stellar luminosity, effective temperature
and distance are from the study of Marchenko et al. (\cite{march10}).
For the shell nebula that surrounds the star, the radii, the expansion velocity,
the kinematic age, the electron density of the ionized gas, the adopted
value of the electron temperature, the measured abundances and masses of dust
and gas are given.

\begin{table}[h]
\caption{Parameters of the WR 124 and its nebula M1-67.}
\label{table:3}
\centering
\begin{tabular}{llc}\hline\hline\\[-0.10in]
Star  & log $L/L_{\odot}$                       &  5.18 $\pm$ 0.2   \\
      & $T_{\mathrm{eff}}$ (K)                  &  35800 $\pm$ 2000 \\
      & $D$ (kpc)                               &  3.35 $\pm$ 0.67  \\
Nebula & \textit{r}$_{\mathrm{in}}$  (pc)        &  0.65            \\
      & \textit{r}$_{\mathrm{out}}$  (pc)       &  1.00             \\
      & ${\rm v}_{\mathrm{exp}}$ (km s$^{-1}$)  &  50 - 150         \\
      & $t_{\mathrm{kin}}$  (10$^{3}$yr)        &  2.3 - 7.0        \\
      & $n_{\mathrm{e}}$ (cm$^{-3}$)            &  600 $\pm$ 180    \\
      & $T_{\mathrm{e}}$  (K)                   &  7000 $\pm$ 1400  \\
      & N/O                                     &  1.0 $\pm$ 0.5    \\
      & C/O                                     &  0.46 $\pm$ 0.27  \\
      & $M_{\mathrm{dust}}$  (M$_{\odot}$)      &   0.22 $\pm$ 0.04 \\
      & $M_{\mathrm{ion. gas}}$  (M$_{\odot}$)  &  0.5 - 3.5        \\
\hline
\end{tabular}
\end{table}

Concerning the expansion velocity of the nebula, Sirianni et
al. (\cite{sir98}) discovered a bipolar outflow with an expansion
velocity of $88$ km s$^{-1}$, apart from a spherical shell that
expands with a velocity of $46$ km s$^{-1}$. Van der Sluys and Lamers
(\cite{sluy03}) found an expansion velocity of $150$ km s$^{-1}$ for
the structure of the nebula that expands freely and which is not
located on the surface of the bow shock. Consequently, the
kinematic age, t$_{\mathrm{kin}}$, of the nebula M1-67 can be
estimated, considering an expansion velocity that can be between $50$ and
$150$ km s$^{-1}$. Since the nebula extends up to $1$ pc from its central
star, its kinematic age is $t_{\mathrm{kin}} = r /{\rm
v}_{\mathrm{exp}}$ = $(6.5$ - $20.0)\times10^3$ years. The temporal
difference between the inner and the outer radius of the nebula is
$(2.3$ - $7.0)\times10^3$ years.

The N/O ratio calculated in this study is the same as the ratio
measured in the nebula around the LBV WRAY 15-751 (Vamvatira-Nakou et
al. \cite{vamv13}) and is similar to the ratio of the nebula around the
LMC LBV R127 (Smith et al. \cite{smith98}). With respect to the solar
N/O abundance ratio used by Ekstr\"{o}m et al. (\cite{eks12})\footnote{These
abundances are derived from Asplund et al. (\cite{asp05}) and differ slightly
from those given in the more recent work of Grevesse et al. (\cite{grev10}).
},
it is enhanced
by a factor of $8$. Considering the 12+log(N/H) abundance of $8.45$
calculated by Esteban et al. (\cite{est91}) and our calculation for
the C/O abundance ratio, we find that these ratios correspond to an
enhancement in N/H by a factor of 5 and a depletion in C/H and O/H by roughly
a factor of 2 with respect to the solar abundances used in the models
of Ekstr\"{o}m et al. (\cite{eks12}). Our results point to a nebula
composed of processed material as is the case of other nebulae around
LBVs and WR stars (Smith \cite{smith97}).

The total gas mass and the corresponding mass-loss rate during the
ejection cannot be estimated without making assumptions. A typical
value for the dust-to-gas ratio is $100$ in such stellar
environments. With this assumption, the total nebular mass is about
$22$ M$_{\odot}$, using the calculated dust mass. But this ratio might
be lower as in the case of the LBV WRAY 15-751 (Vamvatira-Nakou et
al. \cite{vamv13}), where it was calculated to be $40$. In this case,
the mass is about $9$ M$_{\odot}$, still a factor 2 higher than the
sum of the ionized and neutral gas mass, suggesting that the mass of
the gas might be underestimated.

We also compute a lower limit to the total mass-loss rate using the
ionized gas, neutral gas, and dust masses. Since the ionized gas mass
varies with respect to the filling factor used for its calculation
(Sect.~\ref{sec:emission line spectrum}), we use the lower value given
in Table~\ref{table:3}. The lower limit of the mass-loss rate at the
time of the nebular ejection is then calculated to be $\log \dot{M}$ =
$-4.0$, where $\dot{M}$ is in M$_{\odot}$~yr$^{-1}$, considering the
higher value of the temporal difference between the inner and the
outer radii of the nebula (which corresponds to the lower expansion
velocity).

\begin{figure}[t]
\resizebox{\hsize}{!}{\includegraphics*{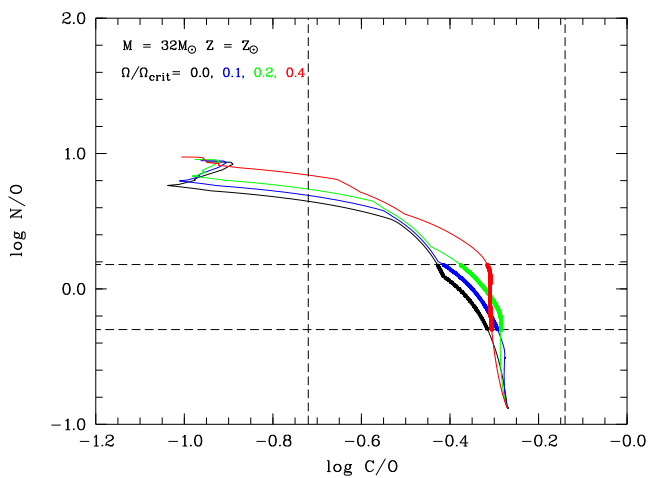}}
\caption{Evolution of the N/O versus the C/O surface abundance
ratios for a 32 M$_{\odot}$ star of solar metallicity and for initial
rotation rates $\Omega/\Omega_{\rm{crit}}$ from 0 to 0.4, using the models
of Ekstr\"{o}m et al. (\cite{eks12}). The dashed lines correspond to the
values measured for the nebula M1-67, with their
errors. The thicker lines emphasize the part of the tracks compatible
with the measurements. For clarity, the tracks are stopped at the
beginning of the blue loop (data point n$^{\rm o}$ 210 in
Ekstr\"{o}m et al. \cite{eks12}).}
\label{m1-67_nso_cso}
\end{figure}

\begin{figure}[t]
\resizebox{\hsize}{!}{\includegraphics*{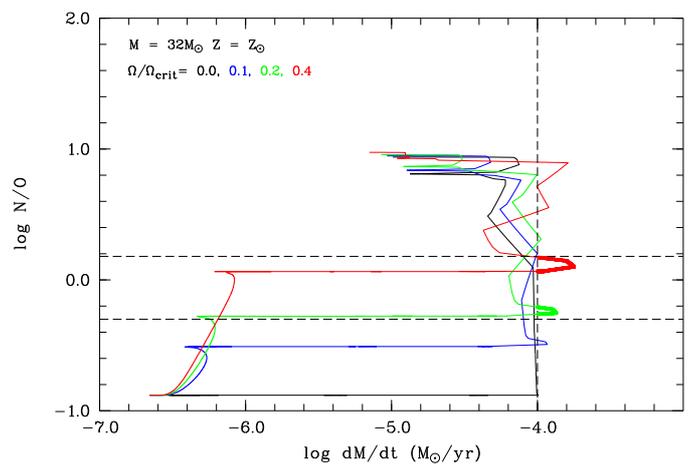}}
\caption
{Evolution of the N/O surface abundance ratio as a function of the
mass-loss rate for a 32 M$_{\odot}$ star of solar metallicity and for
initial rotation rates $\Omega/\Omega_{\rm{crit}}$ from 0 to 0.4,
using the models of Ekstr\"om et al. (\cite{eks12}). The dashed lines
correspond to the calculated value of N/O for the nebula M1-67, with
its errors, and to the lower limit of the mass-loss rate. The thicker
lines emphasize the part of the tracks compatible with the
measurements.  For clarity, the tracks are stopped at the beginning of
the blue loop (data point n$^{\rm o}$ 210 in Ekstr\"om et
al. \cite{eks12}).
}
\label{m1-67_nsomdot}
\end{figure}

\begin{figure}[t]
\resizebox{\hsize}{!}{\includegraphics*{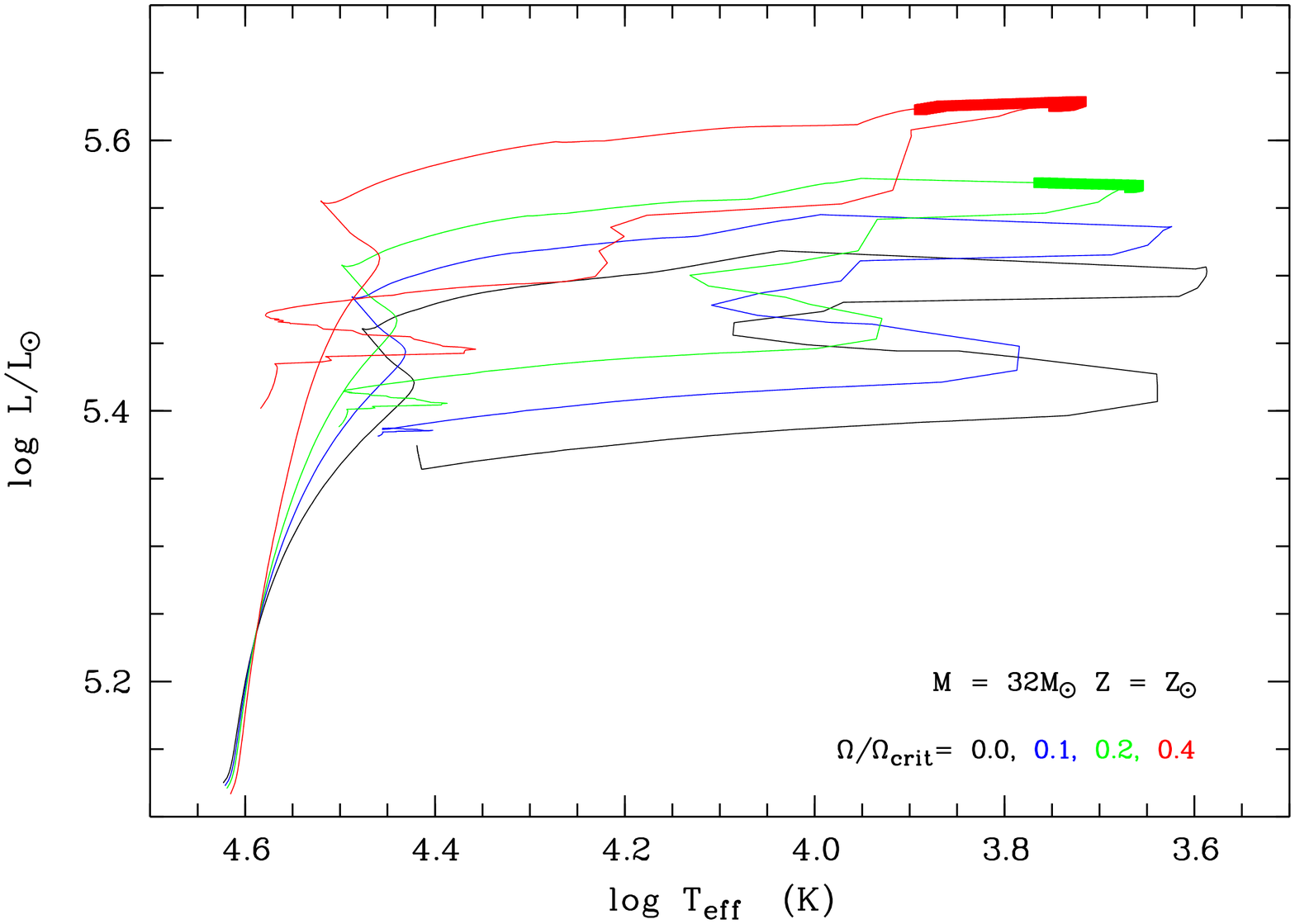}}
\caption{Evolutionary path in the HR diagram of a 32 M$_{\odot}$ star
of solar metallicity and for initial rotation rates
$\Omega/\Omega_{\rm{crit}}$ from 0 to 0.4, using the models of
Ekstr\"om et al.  (\cite{eks12}). The thicker lines emphasize the part
of the tracks compatible with the N/O and C/O abundance ratios and the
mass-loss rate derived from the nebula M1-67. For clarity, the tracks
are stopped when the current values of the stellar parameters
(Table~\ref{table:3}) are reached, considering their errors (data
point n$^{\rm o}$ 310 in Ekstr\"om et al. (\cite{eks12}).}
\label{m1-67_hr}
\end{figure}

These results can be used to constrain the evolutionary phase of the
central star at which the nebula was ejected with the help of the
evolutionary models of Ekstr\"om et al. (\cite{eks12}), keeping in
mind that the models of massive star evolution are very uncertain in
the post-main-sequence evolutionary phases because they do not include
any eruptive event so that the mass-loss rate recipes are not well
known (Smith \cite{smithN_14}).  The limit of the mass-loss rate along
with the calculated nebular abundance ratios of N/O and C/O that are
assumed to be the surface abundance ratios at the time of the nebular
ejection were thus compared to the theoretical evolution of these
parameters according to the models of Ekstr\"om et
al. (\cite{eks12}). The stellar initial mass is constrained by the
stellar luminosity (Table~\ref{table:3}). Models with rotation rates
$\Omega/\Omega_{\rm{crit}} = 0.0 - 0.4$ and solar metallicities were
considered. The evolution of the C/O abundance ratio versus the N/O
abundance ratio for a star of initial mass $32$ M$_{\odot}$ is
illustrated in Fig.~\ref{m1-67_nso_cso}.  The calculated values for
these two parameters with their errors are plotted with dashed lines,
while the part of these tracks that is compatible with our results is
emphasized with thicker lines. In Fig.~\ref{m1-67_nsomdot} the
evolution of the mass-loss rate versus the N/O abundance ratio is
plotted. The constraints from these two diagrams are then reported in
the Hertzsprung-Russel (HR) diagram for a star of $32$ M$_{\odot}$ to
identify which stellar evolution stages correspond to our measurements
(Fig.~\ref{m1-67_hr}).

Models of different initial stellar mass were tested for compatibility
with the calculated abundance ratios and mass-loss rate. The data are
compatible with the evolutionary tracks of a star with an initial mass
between 30M$_{\odot}$ and 35M$_{\odot}$ and solar metallicity, where the
$32$ M$_{\odot}$ model is the most compatible one. High rotational
velocities cannot be excluded. The ejection of the nebula (thick part
of the model tracks of Fig.~\ref{m1-67_hr}) took place when the star
was in a RSG/yellow supergiant (YSG) phase. The star then moved
toward its current WR phase. The evolutionary tracks of this model
are compatible with the current location of the star on the HR diagram
according to its stellar parameters, within the uncertainties. Models
with higher initial mass are not compatible with the current location
of the star because they predict higher luminosity. Models with lower
initial mass are not compatible with the constraints from the
calculated abundance ratios and mass-loss rate.

The time elapsed between the ejection phase and the WR phase, as
computed from the model, is at least two times longer than the
kinematic age of the nebula calculated with the adopted expansion
velocity (Table~\ref{table:3}). This means that the nebula is expected
to be older and bigger. This difference can be explained by the
presence of the bow shock caused by the fast stellar motion. In this
case, the radius of the nebula toward its front is fixed and does not
change with time. So the ejected material from the star flows on the
surface of the bow shock (Van der Sluys and Lamers
\cite{sluy03}). This results in a smaller size of the nebula and a
correspondingly kinematic age lower than the actual age.

%______________________________________________________________
\section{Conclusions}
\label{sec:conclusions}

The \textit{Herschel} photometric and spectroscopic data analysis of
the nebula M1-67 around the Wolf-Rayet star WR 124 has been presented,
together with optical imaging data. The images show a clumpy dusty
nebula that coincides with the gas nebula, indicating that the dust
and the gas are mixed.  As for LBV nebulae, a photodissociation region
was revealed from the infrared spectroscopic data analysis, either
surrounding the ionized gas nebula or originating in the clumps.

The dust nebula model was performed using a two-dimensional radiative
transfer code. Two populations of dust grains with different grain
sizes but with the same composition--olivines with 50/50 Fe to Mg
abundance--were needed to reproduce the infrared SED.  Large dust
grains appear necessary.  Large grains were also found in the dust
nebula around the LBV WRAY 15-751 (Vamvatira-Nakou et
al. \cite{vamv13}), the LBV AG Car (Vamvatira-Nakou et
al. \cite{vamv15}) and the yellow hypergiant Hen 3-1379, which is a
possible pre-LBV (Hutsem\'ekers et al. \cite{hut13}). Models by
Kochanek (\cite{koch11,koch14}) show that large
dust grains can be produced during a LBV eruption.

The analysis of the emission line infrared spectrum points toward a
nebula composed of enriched material, which is a common feature in the
nebulae of WR and LBV that show the CNO-cycle imprints.

The calculated mass-loss rate and the abundance ratios were used to
constrain the evolutionary stage of the star when the nebular
ejection took place, with the help of theoretical stellar evolution
models.  The results suggest that the ejection of the material that
forms the nebula M1-67 around the Wolf-Rayet star WR 124 occurred
during a RSG/YSG phase of a star with initial mass of $32$
M$_{\odot}$.

\begin{acknowledgements}
We thank Dr. Nick Cox for providing the scanamorphos images
and Dr. Ya\"{e}l Naz\'{e} and Prof. Gregor Rauw for their valuable
comments on the manuscript.
C.V.N., D.H., P.R., and M.A.T.G.
acknowledge support from the Belgian Federal Science Policy Office via
the PRODEX Program of ESA. The Li\`ege team also acknowledges
support from the FRS-FNRS (Comm. Fran{\c c}. de Belgique). PACS has
been developed by a consortium of institutes led by MPE (Germany) and
including UVIE (Austria); KU Leuven, CSL, IMEC (Belgium); CEA, LAM
(France); MPIA (Germany); INAF-IFSI/OAA/OAP/OAT, LENS, SISSA (Italy);
IAC (Spain).  This development has been supported by the funding
agencies BMVIT (Austria), ESA-PRODEX (Belgium), CEA/CNES (France), DLR
(Germany), ASI/INAF (Italy), and CICYT/MCYT (Spain). Data presented in
this paper were analyzed using “HIPE”, a joint development by the
Herschel Science Ground Segment Consortium, consisting of ESA, the
NASA Herschel Science Center, and the HIFI, PACS and SPIRE
consortia. This research made use of the NASA/IPAC Infrared
Science Archive, which is operated by the Jet Propulsion Laboratory,
California Institute of Technology.
\end{acknowledgements}

%-------------------------------------------------------------------

\begin{appendix}
\section{Emission line fluxes for each spaxel}
\label{sec:appendix}

\begin{table*}[!]
\caption{Line fluxes in each spaxel. A dash indicates a poor S/N or a non-detection. The spatial configuration corresponds to the footprint of the PACS spectral field of view as displayed in Fig.~\ref{m1-67_sp_foot}.}
\label{table:4}
\centering
\begin{tabular}{c c| c | c | c | c | c }
\hline\hline
Ion               & $\lambda$  & $F\pm\Delta F $ & $F\pm\Delta F $ & $F\pm\Delta F $ & $F\pm\Delta F $ & $F\pm\Delta F $ \\
                  & $(\mu m)$         & (10$^{-15}$ W~m$^{-2}$)    & (10$^{-15}$ W~m$^{-2}$)    & (10$^{-15}$ W~m$^{-2}$)    & (10$^{-15}$ W~m$^{-2}$)    & (10$^{-15}$ W~m$^{-2}$)    \\
\hline\hline
                 &             & $\underline{spaxel\ 4,4}$   & $\underline{spaxel\ 4,3}$  & $\underline{spaxel\ 4,2}$  & $\underline{spaxel\ 4,1}$   & $\underline{spaxel\ 4,0}$  \\
$[\ion{O}{i}]$   & 63     &  -                & 0.047 $\pm$ 0.008 & 0.067 $\pm$ 0.008 &  -                & -                 \\
$[\ion{N}{ii}]$  & 122    & 0.470 $\pm$ 0.026 & 0.551 $\pm$ 0.031 & 0.420 $\pm$ 0.025 & 0.124 $\pm$ 0.010 & 0.170 $\pm$ 0.012 \\
$[\ion{O}{i}]$   & 146    & -                 & -                 &  -                & -                 &  -                \\
$[\ion{C}{ii}]$  & 158    & 0.068 $\pm$ 0.005 & 0.094 $\pm$ 0.006 & 0.066 $\pm$ 0.006 & 0.020 $\pm$ 0.003 & 0.028 $\pm$ 0.003 \\
$[\ion{N}{ii}]$  & 205    & 0.010 $\pm$ 0.002 & 0.020 $\pm$ 0.003 & 0.020 $\pm$ 0.003 & 0.005 $\pm$ 0.002 & -                 \\
\hline
                 &             & $\underline{spaxel\ 3,4}$   & $\underline{spaxel\ 3,3}$  & $\underline{spaxel\ 3,2}$  & $\underline{spaxel\ 3,1}$   & $\underline{spaxel\ 3,0}$  \\
$[\ion{O}{i}]$   & 63     & -                 & 0.160 $\pm$ 0.016 & 0.070 $\pm$ 0.009 &  -                & 0.010 $\pm$ 0.007 \\
$[\ion{N}{ii}]$  & 122    & 0.599 $\pm$ 0.032 & 0.764 $\pm$ 0.040 & 0.451 $\pm$ 0.025 & 0.195 $\pm$ 0.011 & 0.297 $\pm$ 0.018 \\
$[\ion{O}{i}]$   & 146    & -                 & 0.004 $\pm$ 0.001 & -                 & -                 & -                 \\
$[\ion{C}{ii}]$  & 158    & 0.105 $\pm$ 0.008 & 0.128 $\pm$ 0.008 & 0.081 $\pm$ 0.006 & 0.037 $\pm$ 0.003 & 0.050 $\pm$ 0.004 \\
$[\ion{N}{ii}]$  & 205    & 0.019 $\pm$ 0.003 & 0.025 $\pm$ 0.003 & 0.018 $\pm$ 0.002 & 0.006 $\pm$ 0.001 & 0.012 $\pm$ 0.003 \\
\hline
                 &             & $\underline{spaxel\ 2,4}$   & $\underline{spaxel\ 2,3}$  & $\underline{spaxel\ 2,2}$  & $\underline{spaxel\ 2,1}$   & $\underline{spaxel\ 2,0}$  \\
$[\ion{O}{i}]$   & 63     & 0.063 $\pm$ 0.010 & 0.089 $\pm$ 0.008 & 0.148 $\pm$ 0.011 & 0.012 $\pm$ 0.005 & 0.008 $\pm$ 0.005 \\
$[\ion{N}{ii}]$  & 122    & 0.710 $\pm$ 0.038 & 0.769 $\pm$ 0.040 & 0.624 $\pm$ 0.033 & 0.315 $\pm$ 0.018 & 0.316 $\pm$ 0.018 \\
$[\ion{O}{i}]$   & 146    & 0.004 $\pm$ 0.001 & -                 & 0.007 $\pm$ 0.001 & -                 &  -                \\
$[\ion{C}{ii}]$  & 158    & 0.115 $\pm$ 0.007 & 0.139 $\pm$ 0.009 & 0.110 $\pm$ 0.008 & 0.062 $\pm$ 0.005 & 0.062 $\pm$ 0.005 \\
$[\ion{N}{ii}]$  & 205    & 0.024 $\pm$ 0.003 & 0.031 $\pm$ 0.004 & 0.026 $\pm$ 0.003 & 0.018 $\pm$ 0.002 & 0.019 $\pm$ 0.003 \\
\hline
                 &             & $\underline{spaxel\ 1,4}$   & $\underline{spaxel\ 1,3}$  & $\underline{spaxel\ 1,2}$  & $\underline{spaxel\ 1,1}$   & $\underline{spaxel\ 1,0}$  \\
$[\ion{O}{i}]$   & 63     & 0.031 $\pm$ 0.010 & 0.048 $\pm$ 0.007 & 0.076 $\pm$ 0.009 & 0.084 $\pm$ 0.010 & 0.052 $\pm$ 0.007 \\
$[\ion{N}{ii}]$  & 122    & 0.547 $\pm$ 0.031 & 0.613 $\pm$ 0.034 & 0.740 $\pm$ 0.040 & 0.642 $\pm$ 0.035 & 0.745 $\pm$ 0.040 \\
$[\ion{O}{i}]$   & 146    & -                 & -                 & 0.004 $\pm$ 0.002 & 0.005 $\pm$ 0.002 & 0.003 $\pm$ 0.001 \\
$[\ion{C}{ii}]$  & 158    & 0.084 $\pm$ 0.006 & 0.104 $\pm$ 0.007 & 0.118 $\pm$ 0.008 & 0.091 $\pm$ 0.006 & 0.112 $\pm$ 0.007 \\
$[\ion{N}{ii}]$  & 205    & 0.021 $\pm$ 0.004 & 0.023 $\pm$ 0.003 & 0.029 $\pm$ 0.003 & 0.023 $\pm$ 0.003 & 0.020 $\pm$ 0.002 \\
\hline
                 &             & $\underline{spaxel\ 0,4}$   & $\underline{spaxel\ 0,3}$  & $\underline{spaxel\ 0,2}$  & $\underline{spaxel\ 0,1}$   & $\underline{spaxel\ 0,0}$  \\
$[\ion{O}{i}]$   & 63     &  -                & -                 & 0.075 $\pm$ 0.008 & 0.180 $\pm$ 0.015 & 0.063 $\pm$ 0.014 \\
$[\ion{N}{ii}]$  & 122    & 0.279 $\pm$ 0.018 & 0.394 $\pm$ 0.023 & 0.751 $\pm$ 0.041 & 0.848 $\pm$ 0.045 & 0.719 $\pm$ 0.038 \\
$[\ion{O}{i}]$   & 146    & -                 & -                 & -                 & 0.006 $\pm$ 0.001 & 0.006 $\pm$ 0.001 \\
$[\ion{C}{ii}]$  & 158    & 0.042 $\pm$ 0.004 & 0.069 $\pm$ 0.004 & 0.122 $\pm$ 0.009 & 0.139 $\pm$ 0.008 & 0.104 $\pm$ 0.007 \\
$[\ion{N}{ii}]$  & 205    &  -                & 0.019 $\pm$ 0.003 & 0.016 $\pm$ 0.002 & 0.021 $\pm$ 0.002 & 0.018 $\pm$ 0.003 \\
\hline
\end{tabular}
\end{table*}

The emission line flux measurements from the spectrum of
each one of the 25 spaxels are given in Table~\ref{table:4}. The first
column contains the detected ions. Subsequent columns contain
the line fluxes in W/m$^{2}$ along with their errors. In addition,
the spaxel numbers (Fig.~\ref{m1-67_sp_foot}) are given in every
cell of the table. The quoted uncertainties are the sum of the line-fitting
uncertainty plus the uncertainty due to the position of the continuum.

\section{Emission line fluxes from ISO data}
\label{sec:appendix_B}

The LWS observations of M1-67 were taken on day 725 of the ISO mission. They
consist of a 2228-s "on" spectrum (80\arcsec\ circular aperture) and a 1330-s
"off" spectrum acquired 7\arcmin\ away from M1-67. 

The forbidden emission lines $[\ion{N}{ii}]$ 122\ \mbox{$\mu$m} and
$[\ion{C}{ii}]$ 158\ \mbox{$\mu$m} are detected well in this spectrum.
The flux in the $[\ion{N}{ii}]$ 122\ \mbox{$\mu$m} line from the LWS spectrum
is $15.0 \times 10^{-15}$ W~m$^{-2}$, in good agreement with the PACS
measurement of $13.05 \times 10^{-15}$ W~m$^{-2}$ in its $45\arcsec\times45$\arcsec\ IFU.
The LWS on-spectrum gives a flux of $3.8 \times 10^{-15}$ W m$^{-2}$ in the $[\ion{C}{ii}]$
158\ \mbox{$\mu$m} line, while the off-spectrum gives a $[\ion{C}{ii}]$ flux of
$1.45 \times 10^{-15}$ W~m$^{-2}$ from the diffuse ISM, i.e. 38\% of the on-spectrum
value. The net on-off ISO-LWS $[\ion{C}{ii}]$ flux from the nebula is therefore
$2.35 \times 10^{-15}$ W~m$^{-2}$, in agreement with the PACS IFU $[\ion{C}{ii}]$ flux
of $2.13 \times 10^{-15}$ W~m$^{-2}$. The PACS spectrum was obtained in chopping
mode, so that an "off" region spectrum was subtracted from the "on"
spectrum.

\end{appendix}

\end{document}